\documentclass[a4paper,11pt]{article}

\usepackage[utf8]{inputenc}
\usepackage[T1,T2A]{fontenc}
\usepackage{amsmath,amsfonts,amssymb}

 \usepackage{graphicx}
\usepackage{wrapfig}
\usepackage{subfig}
\usepackage{hyperref}
\usepackage[dvipsnames]{xcolor}

\begin{document}

 \begin{titlepage}

\begin{center}
{\LARGE {\bf   Quark, lepton and right-handed  neutrino  production via  inflation }} \\
\vspace{2cm}
{\bf  Duarte Feiteira$^{\;1}$, Fotis Koutroulis$^{\;2,3}$, 
Oleg Lebedev$^{\;1}$ and Stefan Pokorski$^{\;4}$}
\end{center}

\vspace{0.5cm}

\begin{center}
 $^{1}$ \it{Department of Physics and Helsinki Institute of Physics,\\
  Gustaf H\"allstr\"omin katu 2a, FI-00014 Helsinki, Finland}\\
  $^{2}$ {\it Theoretical Physics Division, Institute of High Energy Physics,
Chinese Academy of Sciences, \\19B Yuquan Road,
Shijingshan District, Beijing 100049, China} \\
 $^{3}$ {\it China Center of Advanced Science and Technology, Beijing 100190, China}\\
 $^{4}$ {\it Institute of Theoretical Physics, Faculty of Physics, University of Warsaw, Pasteura 5, \\
 PL 02-093,
Warsaw, Poland}\\
\end{center}

\vspace{2.5cm}

\begin{center} {\bf Abstract} \end{center}
\noindent
 Inflationary expansion of space-time provides us with an efficient particle production mechanism  in the Early Universe.   
The fermion production efficiency depends critically on the particle mass, which
is generated via the Yukawa coupling and sensitive to the corresponding scalar field value.
During inflation, scalar fields experience large quantum fluctuations driving the average field values to the Hubble scale and above.
 This applies, in particular,  to the Higgs field, making the Standard Model fermions very heavy and facilitating their production. 
 Using the Bogolyubov coefficient approach, 
 we compute the corresponding fermion abundance 
 taking into account time dependence of the mass term.  We find that the  Standard Model fermion and the right-handed neutrino production grows dramatically compared to the
 naive estimate based on the low energy masses. The inflationary production mechanism can be the leading source of the right handed neutrinos,
 if they gain a Majorana mass
 from the Yukawa coupling to a light scalar. We also find a lower bound on the mass of fermionic dark matter, which can be produced by inflation.

\end{titlepage}

 \tableofcontents

\section{Introduction}

Particle production in the Early Universe plays an important role in modern cosmology \cite{Mukhanov:2005sc}. First and foremost, it is responsible for reheating, that is, the inflaton energy conversion into the Standard Model (SM) quanta.
This may be accompanied by production of dark matter (DM), either through its coupling to the SM states or via an independent mechanism. The latter possibility has been gaining   traction due to the null direct
DM detection results \cite{LZ:2024zvo}. Dark matter production does not require significant couplings and it can even be produced via gravitational interactions. 

Gravitational particle production    \cite{Parker:1969au}-\cite{Grib:1976pw}    creates an irreducible background in any model of the Early Universe. This is particularly important in the context of non-thermal dark matter  \cite{Lebedev:2022cic}.
More generally, given that   the specifics of reheating remain unknown,  gravitational particle production can make an impact on the Early Universe composition and dynamics.
In particular, it is important to understand the efficiency of the SM particle production due to inflation \cite{Starobinsky:1980te}-\cite{Linde:1981mu}    itself, which so far has not been accomplished in realistic settings.

In this work, we use the Bogolyubov coefficient approach  \cite{Bogolyubov:1958se}  to study 
 fermion production due to  
inflation, 
taking into account time dependence of the fermion masses.
This applies, in particular,  to the SM fermions whose masses are  determined by the Higgs field value. During inflation, the Higgs field is driven to very large values, unless it has a significant positive coupling to the inflaton.
Therefore, the fermions are expected to become heavy, which facilitates their gravitational production. It is known that the abundance ($Y$) of fermions  produced via inflation scales with the (constant) fermion mass $M$ as
   \cite{Chung:2011ck,Koutroulis:2023fgp}
\begin{equation}
Y \propto \left({M\over M_{\rm Pl}}\right)^{3/2}\;,
\label{1}
\end{equation} 
assuming that inflation is followed by a radiation dominated epoch.
The  mass term breaks the conformal symmetry of the fermion action and, as such, plays the critical role in particle production.
In contrast, the conformal symmetry is broken in the scalar sector already in the massless limit, unless the scalar has a specific non-minimal coupling to curvature. This breaking is communicated efficiently to the fermion sector via the Yukawa coupling, 
 making the fermions very heavy in the Early Universe.

During inflation with Hubble rate $H$, a  scalar field average value   tends asymptotically to  \cite{Starobinsky:1986fx,Starobinsky:1994bd}
 \begin{equation}
 \langle s^2 \rangle \rightarrow {3H^4 \over 8\pi^2 m^2} ~~{\rm or}~~\sqrt{3\over 2\pi^2 }\, {\Gamma (3/4) \over \Gamma (1/4) }\, {H^2 \over \sqrt{\lambda}} \;,
 \end{equation}
 depending on which term dominates the scalar potential. Here,
$m\ll H$ is the scalar mass and $\lambda$ is the self-coupling, $V(s) = {1\over 2} m^2 s^2 + {1\over 4} \lambda s^4 \, .$
Hence, unless the scalar is extremely heavy, it takes on a value of order $H$ or above. Since inflation has a  finite duration, the scalar may not reach its asymptotic value, yet its field value will be around $H$ after 60 $e$-folds.
When applied to the Higgs field, this makes the SM fermions up to 11 orders of magnitude heavier than they currently are, and increases dramatically the efficiency of their production.

The above scaling of the fermion abundance   (\ref{1})  only applies to a constant mass term and thus is inadequate for a realistic situation. The Higgs field value drops shortly after inflation, which introduces  the fermion mass time-dependence.
In this work, we compute the fermion abundance with a time-dependent mass using two Ans\"atze  describing a sharp and a slow mass variation in the postinflationary period.  Our results exhibit a different scaling behavior compared to that in (\ref{1}). 
 
Our considerations also apply to production of the right-handed neutrinos via inflation. These  can potentially play the role of dark matter \cite{Boyarsky:2009ix,Boyarsky:2018tvu}, making  the gravitational production channel particularly important. The large neutrino mass in the Early Universe can be generated 
either via the Higgs Yukawa coupling or a coupling to a singlet scalar, which produces a Majorana-type mass term. The latter turns out to be particularly interesting rendering the inflationary neutrino production efficient.
 We also derive the lower bound of about 10 GeV on the fermionic dark matter mass that can be produced by inflation.
It is worth noting that our analysis is based on {\it classical} gravity and as such is well under control.

\section{Basics of  fermion production in an expanding Universe}

Let us briefly summarize basics of fermion production in curved space-time \cite{Parker:1971pt}. Further details can be found in Refs.\,\cite{Chung:2011ck,Koutroulis:2023fgp}. Basics of gravitational particle production 
have been reviewed in \cite{Ford:2021syk,Kolb:2023ydq}.

The Dirac equation in curved space reads
\begin{equation}
(i \gamma^\alpha \nabla_\alpha- M) \,\Psi =0 \;.
\end{equation}
It follows from the action $\int d^4 x \sqrt{|g|} \bar \Psi (i \gamma^\alpha \nabla_\alpha- M) \,\Psi $, where $g_{\mu\nu}$ is the space-time metric, $\nabla$ is the covariant derivative based on a vierbein $e_\alpha^\mu$ and $\alpha$ is the  local Lorentz index.
The Friedmann metric in terms of the conformal time $x_0 \equiv \eta$ is given by
\begin{equation}
ds^2 = a(x_0)^2 \, \eta_{\mu \nu } dx^\mu dx^\nu \;.
\end{equation}
With the help of the Weyl transformation
\begin{equation}
g_{\mu \nu }  = \Omega^2 \tilde g_{\mu \nu } ~,~ \Psi = \Omega^{-3/2} \tilde \Psi ~,~e_\alpha^\mu = \Omega^{-1} \tilde e_\alpha^\mu ~,
\end{equation}
where   $\Omega= a(x_0)$ and   $e_\alpha^\mu$ is the vierbein, 
  $a(x_0)$ can be eliminated from the action, apart from the mass term. 
 Dropping the tilde over the transformed quantities, the  Dirac equation now reads
\begin{equation}
(i \gamma^\mu \partial_\mu-  a(\eta) M) \,\Psi =0 \;.
\end{equation}
It is  the flat space Dirac equation with a time-dependent mass. In general, both $a(\eta)$ and $M$  can evolve in time, leading to particle production.

The solution can be written in terms of the basis functions $U_i,V_i$ with constant coefficients,
\begin{equation}
 \Psi (x) = \sum_i  \left( a_i U_i + b_i^\dagger V_i \right) \;,
 \end{equation}
 where  $i$ denotes collectively the spin and momentum indices.
 In the Heisenberg picture, $a_i,b_i$ are operators with 
  the usual time-independent anti-commutation relations, $\{  a_i , a^\dagger_j\}= \delta_{ij}$, $\{  b_i , b^\dagger_j\}= \delta_{ij}$, etc.
The basis functions are given by
\begin{eqnarray}
 &&U_{{\bf k} ,s} (\eta, {\bf x})= {e^{i {\bf k} \cdot {\bf x}}    \over (2 \pi)^{3/2}} 
 \left(
 \begin{matrix}
  u_{A,k} (\eta) \\
  s \, u_{B,k} (\eta)
 \end{matrix}
 \right)   \otimes h_s ({\bf \hat k}) \;,\\
 &&V_{{\bf k} ,s} (\eta, {\bf x})= -{e^{-i {\bf k} \cdot {\bf x}}    \over (2 \pi)^{3/2}} 
 \left(
 \begin{matrix}
  -u_{B,k}^* (\eta) \\
  s \, u_{A,k}^* (\eta)
 \end{matrix}
 \right) \otimes  h_s (-{\bf \hat k})  \, e^{i \phi}\;,
 \end{eqnarray}
where $k \equiv |{\bf k}|$, ${\bf \hat k} = {\bf k } / |{\bf k}| = (\theta, \phi)$ in spherical coordinates  and  $h_s$ are the helicity 2-spinors satisfying
\begin{equation}
 {\bf \hat k} \cdot \vec{ \sigma} \; h_s = s \, h_s ~~,~~ s= \pm 1 \;,
 \end{equation}
 with $\vec{ \sigma}$ being the Pauli matrices.
 The gamma matrices are taken to be of the form
 \begin{equation}
 \gamma^0 = 
 \left(
 \begin{matrix}
I & 0 \\
 0 & -I
 \end{matrix}
 \right) ~,~ 
 \gamma^i =  \left(
 \begin{matrix}
0& \sigma^i \\
 -\sigma^i & 0
 \end{matrix}
 \right) ~.
 \end{equation}
The 2-spinors satisfy
\begin{equation}
h^{\dagger}_s ({\bf \hat k} ) \,h_r ({\bf \hat k} ) = \delta_{rs} \;,
\end{equation} 
  such that requiring orthonormality of the basis
 \begin{equation}
 (U_i, U_j)=  (V_i, V_j)    =\delta_{ij}~, ~  (U_i, V_j)= 0\;,
 \end{equation}
leads to the condition
  \begin{equation}
|u_A|^2 + |u_B|^2 =1\;.
\label{norm}
 \end{equation}
Here $(f,g) = \int d^3 x \, f^\dagger g$ 
 and  the spacial part of the wave functions is described by the orthonormal set  $ {e^{i {\bf k} \cdot {\bf x}}    \over (2 \pi)^{3/2}} $.

The Dirac equation with the above Ansatz reduces to 
\begin{equation}
 i\partial_\eta        
 \left(
 \begin{matrix}
  u_A \\
  u_B
 \end{matrix}
 \right)
  = \left(
 \begin{matrix}
aM & k \\
 k & -aM
 \end{matrix}
 \right) \; 
 \left(
 \begin{matrix}
  u_A \\
  u_B
 \end{matrix}
 \right)
 ~.~ 
 \label{EOM-u0}
 \end{equation}
It is to be solved with specific boundary conditions. The Bunch-Davies   initial condition   corresponds to the flat space vacuum $a_i | 0 \rangle = b_i | 0 \rangle =0$ \cite{Bunch:1978yq}.
This defines the $in$ wavefunction:
at $\eta\rightarrow -\infty$, the $a(\eta) M$ terms become negligible  and the positive frequency solution is 
\begin{equation}
 \left(
 \begin{matrix}
  u_A \\
  u_B
 \end{matrix}
 \right)^{\rm in}~ \overset{\small{\eta \rightarrow -\infty}}{\longrightarrow} 
 \left(
 \begin{matrix}
  {1/ \sqrt{2}} \\
 {1/ \sqrt{2}}
 \end{matrix}
 \right) \;e^{-i k \eta} \;.
 \end{equation}
 This determines $u_{A,B}$ uniquely.

Another ``vacuum'' can be defined in the infinite future by requiring no particles with respect to the corresponding number operator.
 At $\eta \rightarrow  \infty$, the evolution matrix is diagonal and the positive eigenvalue solution corresponds to
  \begin{equation}
 \left(
 \begin{matrix}
  u_A \\
  u_B
 \end{matrix}
 \right)^{\rm out}~ \overset{\small{\eta \rightarrow \infty}}{\longrightarrow} 
 \left(
 \begin{matrix}
 1\\
0
 \end{matrix}
 \right) \;e^{-i \int \omega(\eta) d\eta} \;,
 \label{eta-inf}
 \end{equation}
with $\omega \rightarrow a(\eta) M$. This asymptotic form, however, does not determine the normalization of $u_B$. To recover it, one expands the evolution matrix in $k/(aM)$ and finds that the 
relevant eigenvector contains $k/(2aM)$ instead of zero in the lower entry.

The solutions with different boundary conditions are related to each other linearly, with constant coefficients.  Under the basis change,
\begin{equation}
\tilde U_{{\bf k} ,s} = \alpha_{{\bf k} ,s} U_{{\bf k} ,s} + \beta_{{\bf k} ,s} V_{-{\bf k} ,s} \;,
\end{equation}
where the tilde refers to the quantity in the new basis. Since $\Psi (x)$ remains invariant, the basis change requires a linear redefinition of the creation and annihilation operators.
Using the orthonormality condition, one finds the Bogolyubov coefficient 
\begin{equation}
\beta_{{\bf k} ,s} = {\rm phase } \times  (u_{A,k} \tilde u_{B,k} -u_{B,k} \tilde u_{A,k}) \;,
\label{Bog}
\end{equation}
where the time-independent  phase  is irrelevant for our purposes. 
Its importance lies in the property that it measures the particle number. 
Identifying the tilded/un-tilded  objects  with $out$/$in$  quantities, one finds that 
$| \beta_{i}|^2 $ corresponds to the average number of particles 
in the $in$ vacuum $|0\rangle$ with respect to the $out$ number operator,
\begin{equation}
 \langle \tilde N_{{\bf k} ,s} \rangle \equiv  \langle 0 | \tilde a_{{\bf k} ,s}^\dagger \tilde a_{{\bf k} ,s} | 0 \rangle = | \beta_{{{\bf k} ,s}}|^2  \;.
 \label{N}
\end{equation}
 The physical number density is then 
 \begin{equation}
 n =\sum_s  \int {d^3 {\bf k} \over (2\pi)^3 a^3} \, |\beta_{{\bf k},s }|^2 \;.
 \label{n0}
 \end{equation}
To compute the Bogolyubov coefficient, one finds  the $in$ and $out$ solutions to the EOM and uses (\ref{Bog}). The solutions are functions of time, while  the 
Bogolyubov coefficient is constant and can be computed at any convenient point,
\begin{equation}
\beta_{{\bf k} ,s}^\prime =0\;,
\end{equation}
where the prime stands for the $\eta-$derivative.
For analytical estimates, one may choose  $\eta \sim \eta_e$ corresponding to the end of inflation. At this point, both $in$ and $out$ solutions can be evaluated reliably.

The above considerations are also valid for a time-dependent mass $M=M(\eta)$ as long as it does not affect the $\eta \rightarrow \pm \infty$ boundary conditions.
In the Standard Model, the fermion mass is determined by the average Higgs field value, which exhibits strong time dependence in the Early Universe.
In particular, it is expected to be very large during and shortly after inflation,  making  fermion production much more efficient.

\section{Fermion production with a time-dependent mass: inflation followed by radiation domination}
 
 The fermion mass in the Early Universe is determined by the environmental effects, e.g. the scalar field expectation value,  and hence is time-dependent. 
 Our starting point is the Dirac equation in curved space-time, where the mass term   $M(\eta)$   can carry explicit time dependence controlled by
 the background evolution.
Denoting the time derivative  $\partial_\eta$ by a prime,  one may  reduce the system   (\ref{EOM-u0})  to the second order differential equations,
  \begin{eqnarray}
  && u_A^{\prime \prime} + \left [i(M a)^\prime +a^2 M^2 +k^2\right] \,u_A=0 \,,\\
  && u_B^{\prime \prime} + \left [-i(M a)^\prime +a^2 M^2 +k^2\right] \,u_B=0 \,,
  \label{uA-eq}
  \end{eqnarray}
  where $k$ is the magnitude of the 3-momentum. Compared to the constant mass case, this system contains an extra term proportional to $M^\prime a$,
  which can have  significant effects.
   For an abrupt mass variation, this term brings in a sharp feature, e.g. a delta-function.
  The wavefunction $u_{A,B} $ remains, however,  smooth.

 The solutions must have certain asymptotic behaviour corresponding to the $in$ or $out$ vacuum. At $a\rightarrow 0$ or $a\rightarrow \infty$, we recover approximately the flat space results with constant $M$,
 which are encoded in the definition of the  $in$ and  $out$ vacua.
 During inflation, the solutions with constant $M$ are the Hankel functions of $\eta$, which also apply to
 an adiabatically changing $M(\eta)$. After inflation, the situation is more complicated and requires a detailed analysis. 
 In what follows, we focus on ``light'' fermions in the sense $M (\eta ) \ll H_e$, where $H_e$ is the Hubble rate at the end of inflation.
 
Our goal is to compute the Bogolyubov coefficient at the end of inflation, $\eta \sim \eta_e$, where both the $in $ and $out$ solutions can be found analytically using reasonable approximations.
We also calculate the Bogolyubov coefficient numerically, without resorting to  approximations and using a smooth transition function $a(\eta)$ between inflation and the radiation domination era. 
Subsequently, we  compute the particle density and the abundance.
 
 We note that some aspects of inflationary fermion production have been studied in \cite{Herring:2020cah,Klaric:2022qly}, although in a different context. Postinflationary perturbative fermion production  via graviton exchange
 was considered in \cite{Barman:2022qgt}, while general gravity-induced operators were analyzed in \cite{Koutroulis:2023fgp}.
 
 \subsection{Fermion masses via Yukawa couplings: the Early Universe }
 
 In the Standard Model, 
 gauge symmetry requires that 
 the fermion masses  $M_f$   be generated via the Yukawa couplings ${1\over \sqrt{2}}\, Y_f h \bar f_L f_R + {\rm h.c.}$, 
   \begin{equation}
M_f= {1\over \sqrt{2}}\,Y_f \langle h \rangle  \;,
\end{equation}
where $h$ is the Higgs field in the unitary gauge and $\langle h \rangle $ is its expectation value.
This also applies to the neutrinos, assuming that they have the right-handed counterparts $\nu_R$,  
   \begin{equation}
 M_\nu^{\rm Dirac}={1\over \sqrt{2}}\,Y_\nu^{} \langle h \rangle \;.
\end{equation}
In addition, the right-handed neutrinos may have a Majorana mass term ${1\over 2} {\cal M} \nu_R \nu_R + {\rm h.c.}$, which 
could in turn  be generated by an  expectation value of an SM singlet $s$. 

The central point of our work is that the Higgs field takes on a large value in the Early Universe\footnote{This assumes the absence of significant (positive) Higgs couplings to the inflaton or the Ricci scalar $R$.}, thereby making the SM fermion very heavy. 
This is required by the de Sitter fluctuations of the scalar fields and is independent of the inflationary model details. 
Within each Hubble patch, the Higgs field takes on an approximately constant value,
\begin{equation}
h(x) = \bar h (x) + {\rm fluctuations} \;,
\end{equation}   
where the classical part $\bar h$ is determined by the long wavelength modes with $k/a \ll H$ and the quantum fluctuations contain high frequency modes. The fermion mass is determined by the average Higgs field value in each Hubble patch,
$\langle h \rangle = \bar h$, with the positive and negative field values yielding  the same   mass by virtue of a chiral rotation.
The size of $\bar h$ is determined by the Starobinsky-Yokoyama probability distribution over all the Hubble patches in the Multiverse \cite{Starobinsky:1994bd},
\begin{equation}
{\bar h}^2 \simeq \langle h^2 \rangle \rightarrow  0.1 {H^2 \over \sqrt{\lambda_h}} \;,
\end{equation}
where $\lambda_h$ is to be evaluated at the inflationary energy scale and we have neglected the bare mass term. 
The above asymptotic value is reached rather quickly, on the timescale of $(\sqrt{\lambda_h} H)^{-1}$.  
Therefore, the fermion mass is determined by  
\begin{equation}
\langle h \rangle   \simeq H_e \;
\end{equation}
in each Hubble patch at the end of inflation.  Here $H_e$ is the corresponding Hubble rate and we
have assumed $\lambda_h (H) \sim 10^{-2}$. One should keep in mind that the Higgs self-coupling at high energies 
is sensitive to the top-quark mass and thus is subject to substantial uncertainties, so the above value should not
be taken as a precise prediction of the Standard Model.

  This shows that the Standard Model fermions become heavy in the Early Universe, by up to 11 orders of magnitude. For  the typical inflationary value $H_e \sim 10^{13}\;$GeV, the bottom quark
  weighs as much as $10^{11}\,$GeV, while the top quark mass is similar to the Hubble rate. The neutrino mass can also be significant, up to 10 GeV, although the result depends on the unknown Yukawa couplings.
  In all cases of interest (apart from the top-quark),
  \begin{equation}
  M_f \ll H_e \;,
  \end{equation}
 such that the fermions can be considered ``light'' in inflationary terms. The top-quark, however, requires a separate consideration.
 
 Gravitational particle production is sensitive to the fermion mass $M$. The fermion abundance produced by inflation itself is given by
  \begin{equation} 
   Y_0\simeq 5\times 10^{-3} \; \left({M\over M_{\rm Pl}}\right)^{3/2} \;,
    \end{equation}
  assuming  radiation-dominated postinflationary dynamics  \cite{Chung:2011ck,Koutroulis:2023fgp}. Hence, for large $\langle h \rangle $ one expects much more efficient particle production than that based on the naive rigid fermion masses.
  This is particularly interesting for the right-handed neutrinos, which can potentially account for dark matter. 
  
  We note that  particle production occurs due to conformal symmetry breaking. In the fermion case, the relevant symmetry breaking parameter is the fermion mass. 
  Although this is a small parameter at low energies, it is driven to large values by the conformally-breaking scalar dynamics in the Early Universe. The latter 
  is significant as long as the non-minimal coupling to gravity takes on a value different from 1/6. Hence, large symmetry breaking effects  in the scalar sector feed into
  the fermion dynamics.     
   
   In reality, the Early Universe fermion masses are not constant and depend on the Higgs or other scalar field dynamics.
   Specifically, the Higgs field goes through the following stages:
   \begin{itemize}
   \item{starting with arbitrary initial conditions, it reaches the value of order $H_e$ during inflation}
   \item{remains constant after inflation until the Hubble rate decreases to the level of the effective Higgs mass}
   \item{starts oscillating in the quartic potential and decays into the SM radiation}
   \item{takes on the electroweak value $v$ at late times}
   \end{itemize}
    Thus, the fermion masses remain large for some time after inflation and then decrease to the standard values. The precise way the masses decrease depends on complicated non-perturbative
    dynamics. To capture its  main features, we consider two extreme possibilities:   an abrupt drop and a slow thermal-like mass evolution. Presumably, the realistic situation is in between the two.
    
    On the other hand, the fermion mass dynamics during inflation is well understood: the mass term evolves  adiabatically  and reaches the terminal value within 10 Hubble times or so. 
    In the distant past, $k/a \gg M_f$, such that the mass does not affect the initial Bunch-Davies state of the fermions\footnote{We may assume the Bunch-Davies fermion vacuum at the beginning of inflation.
    In contrast, for scalars, such an assumption would be inadequate due to possible existence of  a significant scalar condensate  with $k\simeq 0$ \cite{Feiteira:2025rpe}.}.
    We therefore may approximate the fermion wave function at the late stages of inflation 
    by that of a free fermion with fixed mass $M_f= {1\over \sqrt{2}}Y_f \langle h \rangle$, where  $\langle h \rangle \simeq H_e$.

\begin{figure}[h!]
    \centering
    \includegraphics[width=0.5\textwidth]{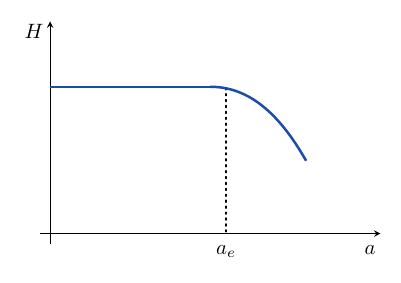}
    \caption{Hubble rate evolution assumed in this work. Inflation ends at $a\sim a_e$ and is followed by the radiation or matter domination epochs.}
    \label{hubble}
\end{figure}

    To compute the particle abundance, we use a smooth transition function $a(\eta)$ from inflation  to the radiation or matter dominated epochs. We neglect small oscillations of the scale factor induced by
    the inflaton oscillations around the minimum.   These can, in principle, have a significant effect on particle production  \cite{Ema:2015dka} which we estimate in  Sec.\,\ref{upper-bound}. In the rest of the paper, we focus on the
    smoothed out or averaged version of the scale factor function. The corresponding Hubble rate evolution is shown schematically in Fig.\,\ref{hubble}.

   \subsection{Radiation dominated era}
   
   Suppose that inflation is followed by the epoch, in which the equation of state of the Universe can be approximated by that of radiation. This can happen due to fast reheating or the 
   inflaton potential being locally 
   $\phi^4 $, as in Higgs inflation \cite{Bezrukov:2007ep}. 
   The scale factor  $a(\eta)$ is chosen such that it describes a smooth transition from inflation at early times to radiation domination \cite{Chung:2011ck},   
 \begin{eqnarray}
 && a(\eta) = \left\{  
 \left(  {1\over a_e H_e}  -\eta  \right)^{-1} H_e^{-1}
  ~~{\rm  for }~~ \eta\leq 0 ~~,~~ a^2_e H_e \left( \eta + {1\over a_e H_e}\right) ~ ~{\rm for}~ \eta>0 \right\}~,\\
 && H(\eta) = \left\{H_e ~~{\rm for}~~ \eta \leq 0 ~~,~~ H_e \, (a_e/a)^2 ~ ~{\rm for}~ ~\eta>0 \right\}~,
 \end{eqnarray}
 where $a_e$ and $H_e$ are the scale factor and the Hubble rate at the end of inflation, respectively; $H(\eta)= a^\prime /a^2$ in terms of the conformal time.
 Here, $a(\eta)$ is continuously differentiable and $a^\prime$ takes on a constant value $a_e^2 H_e$ in the transition region $|\eta| < 1/(a_e H_e)$. We note that only $a$ and $a^\prime$ 
 appear in the EOM, making the wavefunction well-behaved at the end of inflation.
 At $\eta \gg {1\over a_e H_e}$,  a simple scaling holds, 
 $$   \eta \propto a   \;, $$ 
 which simplifies the EOM. In our numerical analysis, we use the full $a(\eta) $ dependence as above. We find that the produced particle number is insensitive to the details of the transition region between inflation and radiation domination, while it is determined primarily by the lifetime of the Higgs condensate.

  The Higgs condensate and the fermion mass stay constant until the Hubble rate reduces to the level of the Higgs effective mass,
  \begin{equation}
H  \simeq \sqrt{3\lambda_h}   \langle h \rangle   \;. 
\label{H-red}
\end{equation} 
At this stage, the condensate starts oscillating in a quartic potential and scaling down in a radiation-like manner. The oscillations produce gauge bosons and other SM states, leading to fast condensate decay, within
${\cal O}(10)$ oscillations \cite{Enqvist:2015sua}. 
We denote the corresponding scale factor by $a_0$ and parametrize the results in terms of 
\begin{equation}
a_0/a_e  \equiv N \sim {\cal O}({\rm few}) \;.
\label{N-definition}
\end{equation} 
At this stage, the fermion mass starts changing fast, possibly dropping substantially. 
In our numerical analysis, we typically take $N=6$ to account for both the Hubble rate reduction (\ref{H-red}) and finite decay time of the condensate, although 
this only gives a ballpark estimate.

In what follows, we consider two possibilities for the $M(\eta) $ dependence. The simplest option is to use the step-function approximation, which corresponds to fast condensate decay and no other significant mass contributions.
The second option is motivated by thermal effects, in which case the mass term decays more slowly, as a power law in $a$.

\subsubsection{Step-function mass term}

To account for a fast drop in the fermion mass, let us take the mass function of the form
$M\, \theta(\eta_0 -\eta) + m\, \theta(\eta -\eta_0) $ with $M\gg m$.  That is,
 \begin{eqnarray}
  -\infty < \eta < \eta_0 & :& ~~ M(\eta )= M \\
  \eta_0 <\eta< \infty &:& ~~ M(\eta )=m\;,
 \end{eqnarray}
where $\eta_0$ corresponds to the scale factor $a_0$ in (\ref{N-definition}).
For this mass function, the inflationary  $in$ solution retains the standard form, while the $out$ solution must be recalculated.
In what follows, we focus on the $out$ wavefunction  and  drop the superscript $out$ for convenience, while  restoring  it when necessary.
\\ \ \\
\underline{{\bf{\textbf{\emph{out}}} wavefunction}, $\eta >\eta_0$}.  At $\eta >\eta_0$, the EOM for $u_A$ is  
 \begin{equation}
  u_A^{\prime \prime} + \left(    k^2 + im a_e^2 H_e + \eta^2 \, m^2 a_e^4 H_e^2  \right)\, u_A =0 \;,
  \label{uA-rad}
 \end{equation}
while the EOM for $u_B$ is obtained by replacing $m \rightarrow -m$.
The boundary condition is 
\begin{equation}
 \left(
 \begin{matrix}
  u_A \\
  u_B
 \end{matrix}
 \right) \overset{\small{\eta \rightarrow \infty}}{\longrightarrow} 
 \left(
 \begin{matrix}
 1\\
{k\over 2am}
 \end{matrix}
 \right) \;e^{-i \int \omega(\eta) d\eta} \;,
 \label{eta-inf1}
 \end{equation}
 with $\omega \rightarrow a(\eta) m$. 
The solution is a parabolic cylinder function $D_\nu (z)$. 
Defining 
\begin{equation}
  C={k^2 \over 2m a_e^2 H_e} \;,
 \end{equation}
we find
\begin{equation}
  u_A^{} (\eta) = e^{-{\pi\over 4} C} D_{-iC} \left( e^{i\pi/4 } \sqrt{2m \over H(\eta)} \right) \times {\rm phase} \;,
  \label{uA-rad}
 \end{equation}
where the  time-dependent  phase is universal for $u_A$ and $u_B$, and thus irrelevant for our purposes.\footnote{This phase is suppressed by $\ln \eta/ \eta^2$   at large $\eta$.}
The $u_B$ part of the wavefunction is 
 \begin{equation}
  u_B^{} (\eta) = \sqrt{C } \,e^{-{\pi\over 4} C + {i\pi \over 4}} D_{-1-iC} \left( e^{i\pi/4 } \sqrt{2m \over H(\eta)} \right) \times {\rm phase} \;,
  \label{uB-rad}
 \end{equation}
 where the  ``phase'' is the same as that in $u_A$. 
 \\ \ \\
\underline{{\bf{\textbf{\emph{out}}} wavefunction}, $\eta \leq \eta_0$}. 
 The EOM is obtained from  (\ref{uA-rad})  by the replacement  $m\rightarrow M$ with the addition of the delta-function term at $\eta_0$. The delta function can be traded for a derivative jump in the boundary
 conditions, which can also be obtained directly from  (\ref{EOM-u0}),
  \begin{equation}
  u'_A\Bigl\vert_- =  u'_A\Bigr\vert_+    -ia_0(M-m)\,u_A (\eta_0) \;,
  \end{equation}
  where ``-'' and ``+'' refer to the limits from below and above $\eta_0$, respectively.   An analogous relation applies to $u_B$ up to $m,M \rightarrow -m,M$.
 Although the solution can be expressed as a combination of the parabolic cylinder functions, it is more physically meaningful to resort to approximations.
 Since $N\sim {\cal O}(\rm few)$ according to (\ref{N-definition}), the $\eta^2$ term at $\eta < \eta_0$ is suppressed by
 \begin{equation}
 N^2 \,{M\over H_e} \ll 1
 \end{equation}
 compared to the constant $M H_e a_e^2 $ term and can be neglected.\footnote{This condition implies that the Hubble rate at the time of the condensate decay $H_0$ is far above the fermion mass $M$.}
 We thus obtain an equation with a constant frequency,
  \begin{equation}
  u_A^{\prime \prime} +  \alpha\; u_A =0 ~~,~~\alpha \simeq k^2 + iM a_e^2 H_e \;,
  \end{equation}
which is solved by 
 \begin{equation}
 u_A = a_1 \, e^{i\sqrt{\alpha} \eta} + a_2 \, e^{-i\sqrt{\alpha} \eta} \;,
 \end{equation}
 with constant $a_i$.
Similarly, 
  \begin{equation}
 u_B = b_1 \, e^{i\sqrt{\alpha^*} \eta} + b_2 \, e^{-i\sqrt{\alpha^*} \eta} \;.
 \end{equation}
The boundary condition on $u_{A,B}$ and their derivatives at $\eta_0$ determines the coefficients $a_i, b_i$.
The Bogolyubov coefficient can be evaluated at $\eta_e$, which, for our purposes, 
we may approximate by $\eta \sim 0$, hence we 
aim at computing 
\begin{equation}
 u_A (0)= a_1+a_2 ~,~  u_B (0)= b_1+b_2\;.
\end{equation}

 The wavefunction is continuous and its value at $\eta=\eta_0$ is determined by the small-mass solution. Since 
 \begin{equation}
 H({\eta_0} )\gg m\;,
 \end{equation}
  we can use the  $z\rightarrow 0$ expansion 
  \begin{equation}
 D_\nu (z) \simeq {\sqrt{\pi}  \, 2^{\nu/2}      \over \Gamma\left(    -{\nu\over 2}   +{1\over 2}\right)} - {\sqrt{\pi}  \, 2^{\nu/2+1/2}      \over \Gamma\left(    -{\nu\over 2}    \right)}\; z \;,
 \end{equation}
together with 
   \begin{equation}
 |\Gamma (it)|= \sqrt{\pi \over t\, \sinh \pi t} ~~,~~ |\Gamma (it+1/2)|= \sqrt{\pi \over  \cosh \pi t}~~,~~
  |\Gamma (it+1)|= \sqrt{\pi t \over  \sinh \pi t}\;,
 \label{Gammas}
  \end{equation}
 valid for a real $t$.
 Up to an overall phase, we have the following expansion in the vicinity of $\eta_0$,
 \begin{eqnarray}
 && u_A^{}(\eta\geq \eta_0)\simeq  e^{- \pi C /4} 2^{-iC/2} \sqrt{\pi} \, \left[     {1\over \Gamma(iC/2 + 1/2)}   - {\sqrt{2} e^{i\pi/4}  \sqrt{2ma_e^2 H_e}  \over   \Gamma (iC/2)} \; \eta   \right]\;,\\
 &&  u_B^{}(\eta\geq \eta_0)\simeq  \sqrt{C} \,e^{- \pi C /4 +i\pi/4} \, 2^{-iC/2-1/2}\, \sqrt{\pi} \, 
  \left[     {1\over \Gamma(iC/2 + 1)}   - {\sqrt{2} e^{i\pi/4}  \sqrt{2ma_e^2 H_e}  \over   \Gamma (iC/2+1/2)} \; \eta   \right]\;. \nonumber
  \end{eqnarray}
These expressions fix the wavefunction value together with its derivative at $\eta_0$.

As is clear from the expression for the frequency squared $\alpha$, there are two distinct regimes for the wavefunction evolution at $\eta<\eta_0$. At 
\begin{equation}
k^2 \ll M a_e^2 H_e \;,
\end{equation}
the evolution is different for $u_A$ and $u_B$, while for large momenta it is universal. 
The non-universality 
is important for the Bogolyubov coefficient, which measures the degree of misalignment between the $in$ and $out$ vectors,
hence it is only significant below the critical momentum value.
In the regime $k^2 \ll M a_e^2 H_e$, we obtain (up to a phase)
 \begin{eqnarray}
 &&   u_A (0)\simeq   e^{- \pi C /4} 2^{-iC/2}  \, \sqrt{\cosh {\pi C\over 2}}         \times  \left(   1 + i {N^2\over 2} \, {M\over H_e}  +          {\cal O} \left(         {k\over a_e H_e}            \right)  \right)\;,    \nonumber \\
 &&   u_B (0)\simeq      e^{- \pi C /4} 2^{-iC/2}  \, \sqrt{\sinh {\pi C\over 2}}         \times     \left(   1 - i {N^2\over 2} \, {M\over H_e}         +          {\cal O} \left(         {k\over a_e H_e}            \right)              \right)\;,
 \label{uAout-final}
  \end{eqnarray}
 where the universal $ {\cal O} \left(         {k\over a_e H_e}            \right)   $ terms cancel in the Bogolyubov coefficient.
 The $C$-dependent  prefactors correspond to $u_{A,B}$ based on mass $m$ and the net result of the mass change at $\eta_0$ amounts to the rotation of $u_A$ and $u_B$ by opposite phases of order 
 $N^2 \, {M\over H_e}$.
 
 At large $k $, the EOM and the derivative jump are  dominated by the universal $k^2$ term, such that the non-universal phase is  suppressed by $M/k$.  The resulting Bogolyubov coefficient would also be suppressed. 
 \\ \ \\
\underline{{\bf{\textbf{\emph{in}}} wavefunction}}. During inflation, the mass term variation is determined by the relaxation time of the Higgs field, $(\sqrt{\lambda_h} H)^{-1}$, which gives a contribution to the EOM suppressed by $\sqrt{\lambda_h} \ll 1$.
Hence, the mass term changes adiabatically and the EOM can be approximated by 
 \begin{equation}
 \eta^2  u_A^{\prime \prime} + \left(    k^2\eta^2  + \left[ {iM \over H_e} + {M^2 \over H_e^2} \right]  \right)\, u_A =0 \;,
  \label{uA-infl}
\end{equation}
where $M$ is taken to be constant at the later stages of inflation and determined by the terminal value of the Higgs field $\langle h \rangle$. This is an adequate approximation for computing the Bogolyubov coefficient at $\eta \sim \eta_e$.
 The EOM for $u_B$ is obtained by replacing $M \rightarrow -M$.

At $\eta \rightarrow -\infty$, the mass term plays no role and we have the usual Bunch-Davies boundary condition
     \begin{equation}
 \left(
 \begin{matrix}
  u_A \\
  u_B
 \end{matrix}
 \right)^{\rm in}~ \overset{\small{\eta \rightarrow -\infty}}{\longrightarrow} 
 \left(
 \begin{matrix}
  {1/ \sqrt{2}} \\
 {1/ \sqrt{2}}
 \end{matrix}
 \right) \;e^{-i k \eta} \;.
 \end{equation}
The solution is given by the Hankel functions,
 \begin{eqnarray}
&& u_A^{in} (a) =  \sqrt{\pi k  \over 4 a H_e}\, 
  e^{ i {\pi \over 2} (1-iM/H_e)}\, H^{(1)}_{{1/ 2} -{iM/H_e}} \left({k\over a H_e}\right) \;,
   \label{uBinf}\\
  && u_B^{in} (a) = \sqrt{\pi k  \over 4 a H_e}\, 
  e^{ i {\pi \over 2} (1+iM/H_e)}\, H^{(1)}_{{1/ 2} +{iM/H_e}} \left({k\over a H_e}\right) \;.
  \label{uAinf}
 \end{eqnarray}
 We are interested in the momentum range  ${k\over a_e H_e} \ll 1$, which can potentially give significant occupation numbers.
 Hence, we may use  
  the small argument expansion
 \begin{equation}
 H_\nu^{(1)} (x) \simeq - {i 2^\nu \Gamma (\nu)    \over \pi } \; x^{-\nu} \;.
 \end{equation}
 At the end of inflation ($a \simeq a_e$), we thus find 
 \begin{eqnarray}
 &&  u_A \simeq {1\over \sqrt{2}}     \times e^{i {M\over H_e} \, \ln {k\over a_e H_e} } \;, \nonumber \\
   &&  u_B \simeq {1\over \sqrt{2}}     \times e^{-i {M\over H_e} \, \ln {k\over a_e H_e} } \;,
   \label{uAin-final}
\end{eqnarray}
where we have neglected terms of order $M/H_e$ not enhanced by any additional factors.
 \\ \ \\
\underline{{\bf   Particle number}}. The Bogolyubov coefficient 
$|\beta_{{\bf k} }| = |u_{A,k}^{in} u_{B,k}^{out} -u_{B,k}^{in}  u_{A,k}^{out}| \;,$
can be evaluated at $\eta \sim \eta_e$, where our $in$ and $out$ solutions give a good approximation
to the true wavefunctions. Using (\ref{uAout-final}) and (\ref{uAin-final}), we find that 
at $C\ll 1$, the $out$ wavefunction is proportional to $(1,0)^T$, while the $in$ wavefunction has the form
$(1/\sqrt{2},\, 1/\sqrt{2})^T$ up to small corrections. Hence, the  Bogolyubov coefficient  is close to $1/ \sqrt{2}$.
For larger momenta, $C >1$, but still below 
  $\sqrt{Ma_e^2H_e}$, both $in$ and $out$ wavefunctions are close to  $(1/\sqrt{2}, 1/\sqrt{2})^T$, yet the cancellation
  in the Bogolyubov coefficient is incomplete and the leading order result is
  \begin{equation}
  |\beta_{{ k} }|   \simeq  {1\over 2} {M\over H_e}\, \left\vert     N^2 - 2\ln    {k\over a_e H_e}   \right\vert \;.
 \end{equation}
For $N\sim 6$, the $\ln k$ term can be neglected and $ |\beta_{{ k} }| $ is approximately constant in this momentum window.
At yet larger $k \gtrsim \sqrt{Ma_e^2H_e}$, the constant term disappears and $ |\beta_{{ k} }| $ drops further, approaching zero
at very large momenta.

\begin{figure}[h!]
    \centering
    \includegraphics[width=0.60\textwidth]{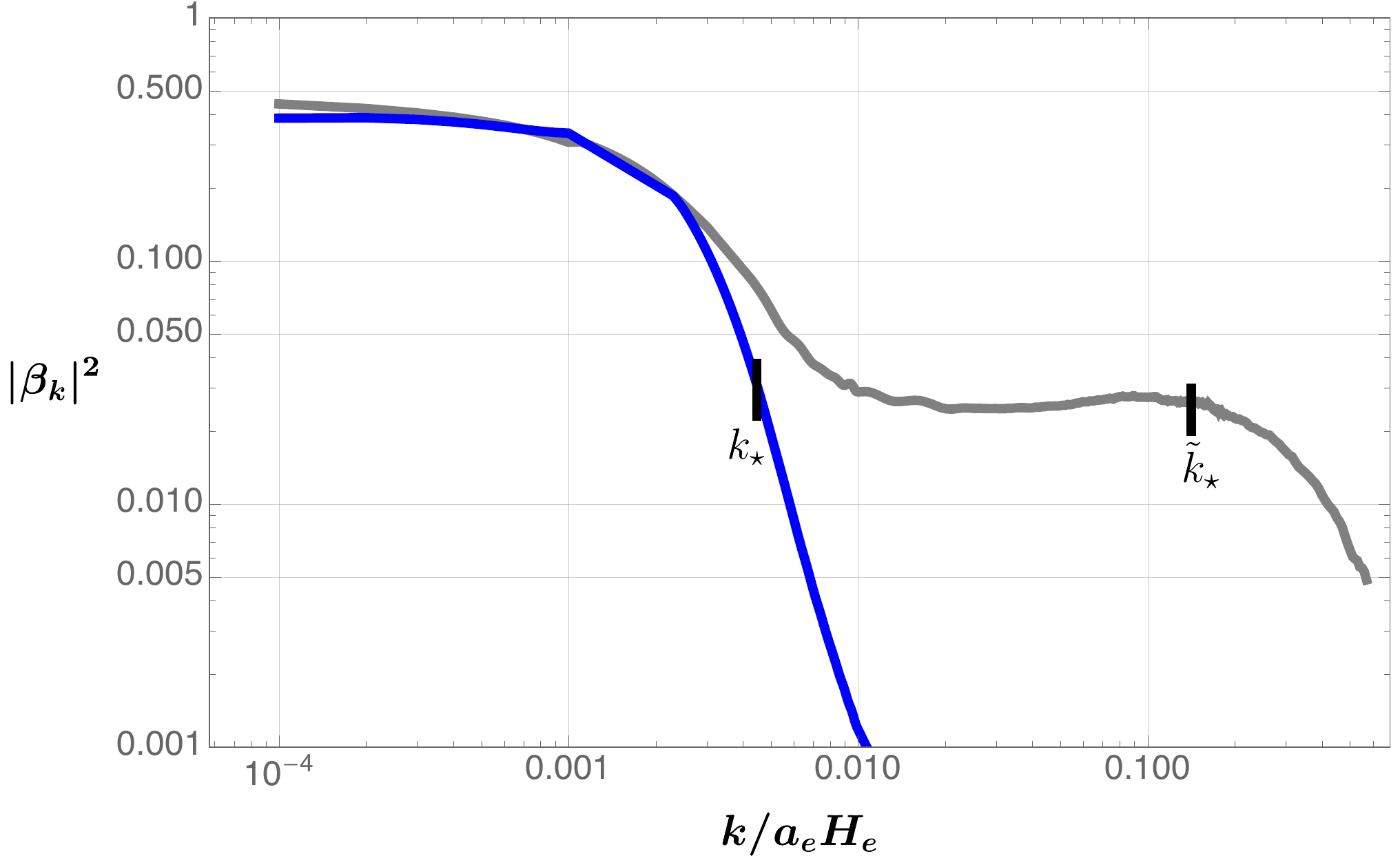}
    \caption{Bogolyubov coefficient squared for   a constant fermion mass (blue) and piece-wise constant mass (grey). Radiation domination after inflation is assumed
    and $m= 10^{-5} H_e\,,\, M= 10^{-2} H_e\,, N=6\,.$ }
    \label{bog-plot}
\end{figure}

Our results are summarized as: 
\begin{eqnarray}
 k\ll k_* &~: ~~&  |\beta_{{ k} }| \simeq {1\over \sqrt{2}} \;,  \nonumber \\
       k_*    \lesssim  k \ll \tilde k_* &~: ~~&  |\beta_{{ k} }|   \simeq  {1\over 2} \,N^2\,{M\over H_e}\;, \nonumber \\
        \tilde k_* \lesssim k &~: ~~&  |\beta_{{ k} }|   \simeq 0 \;,
\end{eqnarray}
with 
\begin{equation}
k_* = \sqrt{2ma_e^2H_e}~~,~~ \tilde k_* = \sqrt{2Ma_e^2H_e} \;,
\end{equation}
where we have defined $\tilde k_*$ with a factor of $\sqrt{2}$ for uniformity of notation. Fig.\,\ref{bog-plot} shows
representative results of 
 our numerical analysis, which does not resort to the approximations made above.
 One clearly sees the step-like features in $|\beta_k|^2$ with the appropriate momentum cutoffs, 
 as expected from our analytical estimates.

The particle number depends on the momentum cutoff cubed. Thus, for a large hierarchy between $m $ and $M$, the main contribution to
the particle density comes from the momentum range between $k_*$ and $\tilde k_*$, even though the corresponding average occupation number 
is small.
In this case, we may approximate the particle density produced via inflation by
  \begin{equation}
 n \sim 4\times  \int {d^3 {\bf k} \over (2\pi)^3 a^3} \,     \theta(\tilde k_* - k)    \,  |\beta_{{ k} }|^2 \;,
 \label{n1}
 \end{equation}
 where $|\beta_{{ k} }|$ corresponds to the momentum range $k_* < k<\tilde k_*$ and the factor of 4 comes from the spin d.o.f. of a Dirac field.
 Thus,
  \begin{equation} 
  n\sim {2\over 3 \pi^2}\,  \tilde k_*^3\, {1\over a^3}\, |\beta_k|^2  = {\sqrt{2} \over 3\pi^2} N^4 \, {M^{7/2} \over H_e^{1/2}} \, {a_e^3 \over a^3}\;.
 \label{n2}
 \end{equation}
  This result is dominated by the high momentum modes up to $\tilde k_*$, which have a low average occupation number of order $N^4 \,M^2/H_e^2$. Note that this 
  quantity can be expressed as $M^2/H_0^2$, where $H_0$ is the Hubble rate at the time of the condensate decay.
  
  The particle abundance is computed according to 
  \begin{equation}
   Y = {n \over s_{\rm SM}} ~~, ~~ s_{\rm SM}={2\pi^2 g_*\over 45}\, T^3\;,
   \end{equation}
 where $s_{\rm SM}$ is the entropy density of the SM thermal bath with temperature $T$ and $g_*$ 
   is the effective number of degrees of freedom in the SM bath.
   This is a meaningful quantity if the particle number is conserved after the production process has completed. In our case, particle production 
   is independent of $m$ and 
   occurs at the early postinflationary stage, not far from the condensate decay time,
   hence $Y$ is conserved after that.  The result is independent of the reheating temperature  as long as the Universe is radiation-like, which could be due to the inflaton oscillations in the $\phi^4$ potential. 
    We find
   \begin{equation} 
 {  Y\sim   10^{-3} \times  N^4  {M^{7/2}    \over M_{\rm Pl}^{3/2}    H_e^2 }}\;.
   \label{Y-M}
    \end{equation}
This gives the abundance of gravitationally produced particles in the regime $m\ll M \ll H_e$ in the case of a short-lived condensate, $N^2 M/H_e \ll 1$, which decays abruptly. Here $N\geq 1$ controls the lifetime of the scalar condensate 
and for the Higgs field one expects $N\sim {\cal O}({\rm few})$.

In the case of the Higgs-induced masses, one may rewrite this result in a more palatable form. 
For instantaneous condensate decay at $a=a_0$, we have 
 $H_e/N^2 = H_0\sim \sqrt{3\lambda_h} \langle h\rangle  \simeq \sqrt{3\lambda_h}  H_e $. Since $M\sim Y_f H_e$, 
 the fermion abundance can be estimated by
 \begin{equation} 
 Y^{\rm SM}\sim   10^{-3} \times  {Y_f^{7/2} \over \lambda_h} \, \left( {H_e    \over M_{\rm Pl}} \right)^{3/2}  \;,
   \label{Y-M-SM}
    \end{equation}
 up to the color multiplicity factor.
 For large $H_e$, this result is enhanced by many orders of magnitude compared to the naive estimate  based on the electroweak fermion masses, $Y^{\rm SM} \sim 10^{-26} \,Y_f^{3/2}$.
 In particular, with $H_e \sim 10^{13}\,$GeV, our result is larger by a factor of $10^{19} \,Y_f^2\,$.
 
 One should keep in mind that our considerations apply to  fermion   production due to the  time-dependent  background. There are, of course, other, more powerful, sources of the SM fermions such as    the
 direct inflaton decay, etc. These create the thermal bath that enters into the $Y$ calculation. While for the SM fermions our analysis does not affect  the conventional approach to reheating\footnote{Inflationary SM fermion production can be relevant to models where the inflaton decays exclusively to the hidden sector states.}, inflationary particle production can be the leading source 
 for very weakly interacting fermions such as the right-handed neutrinos.

 \subsubsection{Slow  effective mass decrease and thermal effects}
  \label{thermal-effects}

  The transition from a large mass $M$ to a small mass $m$ in reality is expected to be  smooth rather than abrupt.
  This can be  due to generation of the effective fermion 
   mass by ``medium'' effects, for example,  via interaction with  the thermal bath.
   The thermal mass may  be    very large soon after inflation but  vanish at late times.
 
 In what follows, we $model$ the transition to the small mass regime using  thermal effects as a template.  Instead of performing the thermal QFT analysis in curved space, 
 we parametrize the time dependence of the effective fermion mass 
   in accordance with the thermal QFT expectations. This should capture the main features of fermion production in a more realistic setting.
 
We treat the thermal bath as the ``medium'' which creates an effective mass due to the coupling to the gauge bosons,
 but {\it does not directly produce fermions} itself.
    In this approach, particle production occurs due to the background and mass time dependence in the Dirac equation, which differs from the direct particle production via, for example, the inflaton decay.  
   Strictly speaking, of course, the gauge and gravitational effects are entangled
 in this system. Yet, our  analysis is helpful and 
 our core results apply more generally, beyond the thermal mass approximation.

    \begin{figure}[h!]
    \centering
    \includegraphics[width=0.60\textwidth]{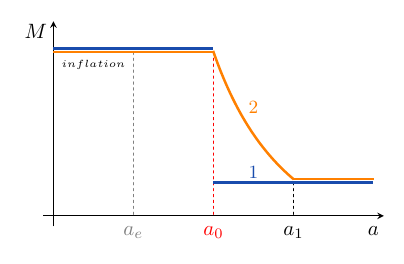}
    \caption{Effective fermion mass evolution: (1) abrupt, (2) smoothed by thermal effects.   }
    \label{M-plot}
\end{figure}

  In what follows, we compute the Bogolyubov coefficient based on a continuous $M(a)$ function
with $$M \propto T \propto 1/a \;, $$
   as long as the thermal mass dominates. 
   Here $T$ is the SM bath temperature and the above scaling assumes fast reheating, i.e. the SM thermal bath takes over the energy balance around the condensate decay time or even earlier.
   We model the $M(\eta)$ function with 3  distinct periods 
     (Fig.\,\ref{M-plot}):
   \\ \ 
   
   (1) $\eta < \eta_0 :$ large constant mass $M$
   
   (2)   $\eta_0 < \eta < \eta_1 :$ thermal mass $M(\eta) \propto 1/a $
   
   (3) $ \eta_1< \eta  :$ small constant mass $m$
  \\ \ \\
  The next step is to compute the $out$ solution by solving the EOM in each period and matching them at the boundaries. The $in$ solution remains the same as before, hence
  we focus entirely on the $out$ wavefunction and drop the $out$ superscript. Since the $\eta\rightarrow \infty $ behavior remains unchanged, it is easiest to start with the late time period.
  \\ \ \\
  \underline{{\bf Late times}.}     In the  regime $\eta>  \eta_1  $, the solutions 
   (\ref{uA-rad}) and  (\ref{uB-rad}) apply. 
  At late times, $H(\eta) \sim T^2/M_{\rm Pl} \ll m$, hence one can use the large argument expansion of the parabolic cylinder function, 
  $$  D_\nu (z) \simeq e^{-z^2/4} \, z^\nu \;,$$
   for $z\gg 1$,
   yielding 
  \begin{eqnarray}
  && u_A (\eta) \simeq e^{-i {m\over 2H(\eta)}}   \, ,\nonumber\\
  && u_B (\eta) \simeq {k\over 2m\,a (\eta)}\, e^{-i {m\over 2H(\eta)}}  \;,
  \end{eqnarray}
  neglecting the sub-leading phase contributions in the limit
  $m\gg H(\eta)$.\footnote{Note also that the positive energy  
    asymptotic     $out$  states are       obtained as an expansion in $k/(a\,m)$. }
    In terms of the conformal time, the Hubble rate is given by $ H = {1\over a_e^2 H_e \eta^2}\,$.
  
  This constant mass regime extends to the earlier times to the point where the thermal mass contribution
  becomes comparable to the bare mass,
  \begin{equation}
  m \sim gT \;,
  \end{equation}
  where $g$ denotes a generic gauge coupling responsible for the thermal mass. In the Standard Model, the Higgs-induced mass also changes due to the temperature dependence of the Higgs potential.
  In our context, however, this effect is unimportant.
     \\ \ \\
   \underline{{\bf Intermediate regime}}. For $\eta < \eta_1 $, the thermal mass becomes more important than the bare mass.  
  Since $T \propto 1/a$ in the SM radiation dominated Universe, 
  \begin{equation}
  a\,M(a) \simeq {\rm const} \;. 
  \end{equation}
  This applies to the period between $a_0$ and $a_1$, hence $a_0 \,M \simeq a_1 \,m$.
  The EOM (\ref{uA-eq}) take the form 
  \begin{eqnarray}
  && u_A^{\prime \prime} + \left(a^2 M(a)^2 +k^2\right) \,u_A=0 \,\\
  && u_B^{\prime \prime} + \left(a^2 M(a)^2 +k^2\right) \,u_B=0 \,.
  \label{uA-eq-1}
  \end{eqnarray}
The equations  are identical for $u_A$ and $u_B$, and contain a constant frequency
 $$   \omega = \sqrt{ a^2 M^2 +k^2}   \;.$$ 
  The solutions are 
  \begin{eqnarray}
  &&  u_A =  a_1 \, e^{i\omega \eta} +  a_2 \, e^{-i\omega \eta} \;, \nonumber\\
   &&   u_B =  b_1 \, e^{i\omega \eta} +  b_2 \, e^{-i\omega \eta} \;,
  \end{eqnarray}
  with constant $a_i,b_i$.
  
  Consider the regime 
  \begin{equation}
  k \ll a_1 m  \;,
  \end{equation}
  Let us first evaluate 
    $u_B$. Matching the wavefunction value and its derivative at $\eta_1$, one finds $b_1/b_2 = {\cal O}(k^2/(a_1^2 m^2))$ and
    \begin{equation}
   u_B  (\eta<\eta_1) \simeq {k\over 2a_1 m}\, e^{-i\omega \eta} \times {\rm const~phase} \;.
  \end{equation}
  Similarly,
   \begin{equation}
   u_A  (\eta<\eta_1) \simeq {  e^{-i\omega \eta} \times {\rm const~phase}} \;.
 \end{equation}
  Therefore, at $\eta_0$ the wavefunction takes the form
  \begin{equation}
  \left(
   \begin{matrix}
  u_A \\
  u_B
 \end{matrix}
 \right)^{} (\eta_0) \sim 
 \left(
 \begin{matrix}
 1\\
 0
 \end{matrix}
 \right) \times {\rm phase}\;,
 \end{equation}
up to corrections of order ${k\over a_1 m}$. Thus, it retains its $\eta \rightarrow +\infty$ asymptotic form.  
  This is a feature of the constant frequency evolution in the period   $\eta_0 < \eta < \eta_1 $.           
  
\begin{figure}[h!]
    \centering
    \includegraphics[width=0.495\textwidth]{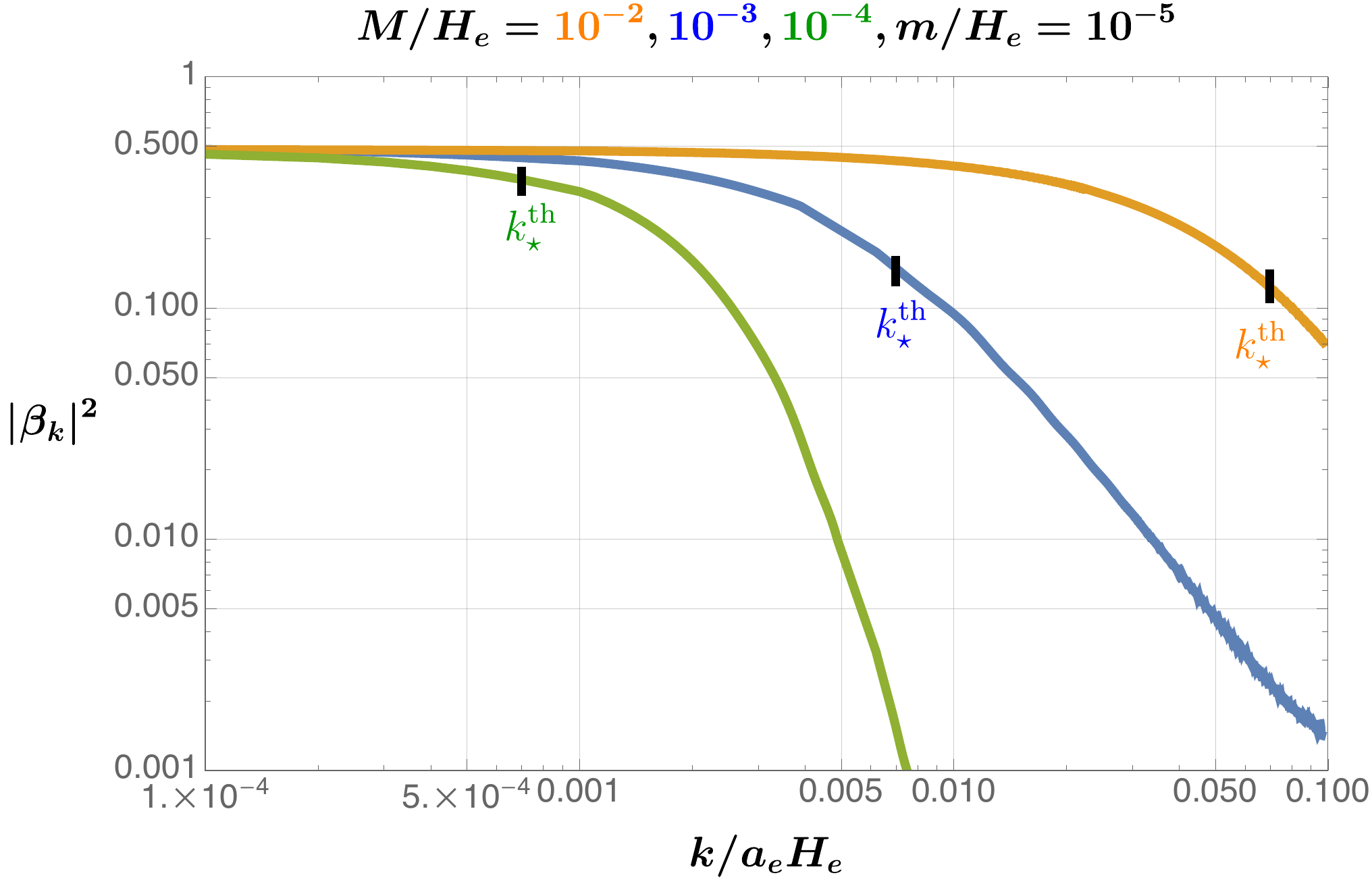}
     \includegraphics[width=0.495\textwidth]{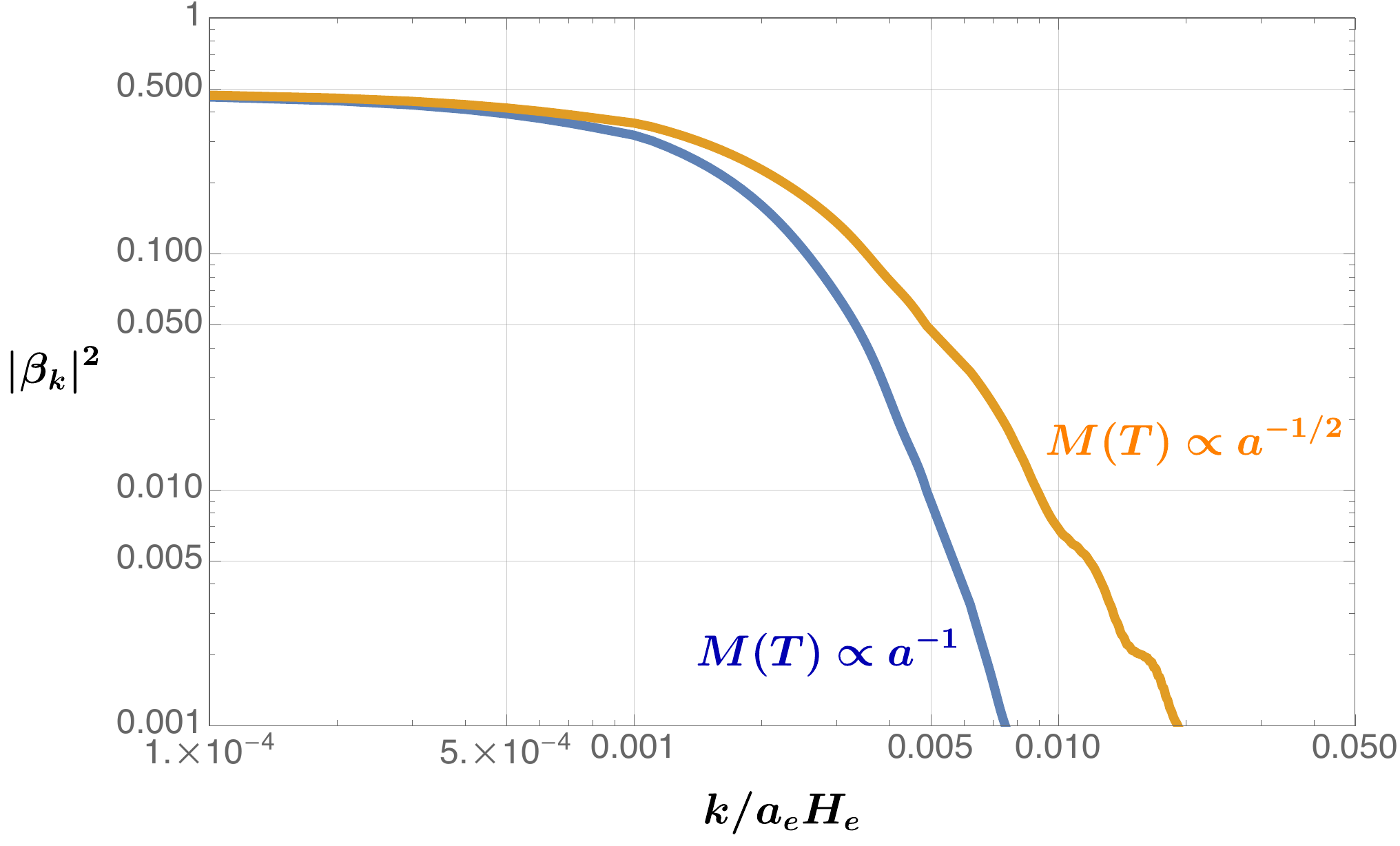}
    \caption{ Bogolyubov coefficient squared for   a fermion with a thermal mass during radiation domination. {\it Left: } The orange, blue, green curves correspond to
       $M/H_e= 10^{-2}, 10^{-3}, 10^{-4} \;,$  respectively, and $m/H_e = 10^{-5}$.  The occupation number drops above $k_*^{\rm th}$. 
       {\it Right:} Comparison of the results for different thermal mass scaling laws: $M(T)\propto a^{-1}$ vs  $M(T)\propto a^{-1/2 }$, with the other parameters fixed.}
    \label{M-thermal-plot}
\end{figure}

  As $k$ approaches $a_1 m$, the wavefunction initial condition at $\eta_1$ tends to the democratic form close to 
  $(1/\sqrt{2}, 1/\sqrt{2})^T$. This form is retained by the constant frequency evolution from $\eta_1$ to $\eta_0$. Thus 
  $k \sim a_1 m=a_0 M $ represents the momentum cutoff for our considerations and, for larger momenta, there are significant cancellations
  in the Bogolyubov coefficient.
  \\ \ \\
\underline{{\bf Early times}}.  
The evolution from $\eta_0$ to $\eta_e $  proceeds as before. It amounts to a small  rotation of the wavefunction by a phase of order  $N^2\, M/H_e$.
This is a subleading effect in the present case and we may neglect it. 
\\ \ \\
\underline{{\bf Particle number}}.  
The $in$ wavefunction remains of the form $(1/\sqrt{2}, 1/\sqrt{2})^T$ for the entire momentum range of interest.  The $out$ wavefunction has the form 
$(1, 0)^T$ for the momenta below $a_0M$, hence 
we may approximate
\begin{eqnarray}
    k  \lesssim k_*^{\rm th}  &~:~~&  |\beta_k| \simeq 1/\sqrt{2}  \;, \nonumber\\
      k> k_*^{\rm th} &~:~~&  |\beta_k| \simeq 0 \;,
\end{eqnarray}
where  the ``thermal'' cutoff is
\begin{equation}
k_*^{\rm th} = a_0 \,M\;.
\end{equation} 
Our numerical results are shown in Fig.\;\ref{M-thermal-plot}, left panel.  They exhibit good agreement with the analytical estimates, in particular, in terms of the position of the momentum cut-off.

Using the $\theta (k_*^{\rm th}-k)$  approximation in the Bogolyubov coefficient,   we find
\begin{equation}
 n \sim  {1\over 3 \pi^2 } N^3 M^3 \, {a_e^3\over  a^3}\;,
 \label{density-thermal}
 \end{equation}
  and 
  \begin{equation}
 Y  \sim  10^{-3} \times N^3 \, {M^3 \over (M_{\rm Pl} H_e)^{3/2}} \;.
 \label{Y-thermal}
 \end{equation}
This is larger than our previous result by the factor $(N \sqrt{M/H_e})^{-1}$, so we conclude that 
$$   Y_{\rm fast}   \sim Y_{\rm slow} \times N \sqrt{M\over H_e} \ll  Y_{\rm slow}\;, $$
where the subscripts ``slow'' and ``fast'' refer to the slow and fast fermion mass variation after the scalar condensate decay.

So far  we have worked in the approximation that the thermal mass is similar to the Higgs-induced mass. However, our results apply more generally.
Let us consider some variations of our assumptions.

\begin{itemize}
\item{   {\it different mass scaling.}   If the SM thermal bath is subdominant in the energy balance, its energy density scales as $1/a^2$ and the corresponding temperature scales as $1/a^{1/2}$ (see, e.g. \cite{Cosme:2024ndc}).
This changes the fermion mass scaling to $M(a)\propto a^{-1/2} \,.$ We find that the  resulting effect on the Bogolyubov coefficient is insignificant, in particular, the momentum cutoff remains almost the same  
 (Fig.\;\ref{M-thermal-plot}, right panel).  }
\item{    {\it sharp features  in $M(\eta)$. }   At early times $\eta \sim {\cal O}(\eta_0)$, the effective fermion mass can change abruptly due to fast non-perturbative effects. Also, the thermal mass could grow  larger than that induced by the Higgs
condensate. To understand the influence of such effects, consider  an abrupt fermion mass change:
$$
      M(\eta) = M \, \theta(\eta_0 - \eta)  + {\cal M} \, \theta (\eta-\eta_0)    ~~,~~  M \ll {\cal M} \ll H_e \;.
$$
The wavefunction analysis is completely analogous to our earlier calculations  with the result
 $$    u_A (0) = u_A (\eta_0) \; \left[ 1  + {\cal O} \left({{\cal M} \over H_e} N^2\right) \right] \;, $$ 
 omitting the irrelevant terms for computing the Bogolyubov coefficient. 
 Therefore, the correction is small as long as the largest mass is significantly below the Hubble rate, $N\, \sqrt{ {\cal M} /H_e} \ll 1$.  
  The Bogolyubov coefficient is determined primarily by the ``thermal'' mass, which decreases slowly over a long period, rather than
  abrupt mass variations at $\eta \lesssim \eta_0$. This is also confirmed by our numerical analysis.
}
\end{itemize}

Our conclusion is that the above result is quite stable to variations of the thermal mass scaling and introduction of sharp features into the mass function.
Note also a significant degree of flavor universality of the result: although the scale $a_1(m_f)$ is flavor dependent and  defined by $g\,T (a_1) \sim m_f$,
the combination $a_1(m_f) \,m_f=a_0 \,M$ is flavor independent and it is this product that determines the abundance.

\section{Standard Model fermion and right-handed neutrino production via inflation}

In this section, we discuss inflationary production of the SM fermions as well as right-handed neutrinos. Classical gravity is responsible for production of the SM states irrespectively of the inflaton couplings,
which is an omnipresent  effect in standard cosmology. For the right-handed neutrinos $\nu_R$, this production channel can be particularly important
given that their couplings could  be 
  very small. Since $\nu_R$
   or, more generally, singlet fermions, may constitute dark matter, the question of inflationary particle production becomes all the more pressing. 
In what follows, we assume that inflation is followed by the radiation domination era, while the matter domination option will be discussed 
in Section \ref{matter}.

 \subsection{Quark and lepton production}

Our results apply directly to quark and lepton production via inflation.
  The Higgs condensate creates a large effective fermion mass which facilitates gravitational particle production. Its efficiency is sensitive
  to the post-inflationary wavefunction evolution. 
  As is clear from the above discussion, one distinguishes two possibilities:  in the first case,
the effective mass drops abruptly to some small value, whereas in the second case, such a decrease happens over a long period,
which we model by the thermal mass contribution. While the postinflationary  dynamics remain largely unknown, a realistic situation
is likely to fall in between these special cases.

  \subsubsection{Sharp effective mass decrease}

  This  possibility corresponds to  fast Higgs condensate decay,   in the absence of other significant sources for the effective fermion mass. This is the case when the thermal or non-equilibrium effects
  on the fermion propagation can be neglected.
  
  The production efficiency is flavor-dependent and controlled by the fermion Yukawa coupling ${ Y}_f$. Since $M_f = {1\over \sqrt{2}} { Y}_f  \langle h \rangle $, for $ \langle h \rangle \simeq H_e$ and a step-function mass profile, we get
  \begin{equation}
n_f^{\rm fast} \sim {\cal O}(10) \times  N_c \, { Y}_f^{7/2} \,H_e^3 \, {a_e^3\over a^3} \;,
\label{nf-step}
\end{equation}
 where we have taken $N=6$ and $N_c$ is the color multiplicity. 
 The resulting abundance   (\ref{Y-M})    of heavier fermions is much larger than that for lighter fermions, as long as $M_f \ll H_e$.
  The production is most efficient at early times, not far from the 
  Higgs condensate decay time.
  
 It is instructive to compare this result to the particle density $n_f (m_f)$   computed with constant low energy fermion masses $m_f = {1\over \sqrt{2}} Y_f \,v$. The 
 ratio of the two is
 \begin{equation}
{ n_f^{\rm fast} \over  n_f (m_f)} \sim \left(  { M_f \over m_f}  \right)^{3/2} \times N^4 \, {M_f^2 \over H_e^2} \sim 10^{19} \, Y_f^2 \;,
 \end{equation}
 for $H_e \sim 10^{13}\,$GeV.
 We observe that our result exceeds $n_f (m_f)$  by many orders of magnitude, e.g. 15 in the case of the bottom quark. 
  For very  light fermions, the discrepancy becomes smaller due to 
  the suppression factor $Y_f^2$, which accounts for 
 small average occupation numbers in the regime $N \sqrt{M_f / H_e} \ll 1$.

   The energy density  of the produced particles is  
   dominated by the bottom quarks and 
  remains  tiny compared to that of the inflaton, $\rho_\phi \sim H^2 M_{\rm Pl}^2$. Indeed, one can estimate or, more precisely, bound  the characteristic energy   of the fermion  quantum by $\tilde k_*/a_0$,
   which yields 
\begin{equation}
  \rho_{f} \lesssim   { Y}_f^{4} \,H_e^4\sim M_f^4 \, 
  \end{equation}
  around  the time of the condensate decay. 
  Clearly, this contribution is  unimportant for reheating, but can be significant in other contexts.

   \subsubsection{Slow effective mass decrease}
  
  This, presumably more realistic, option corresponds to the presence of significant thermal or non-equilibrium contributions to the fermion mass after the condensate has decayed. We model the effective mass decrease by the time-dependent thermal mass,
  although the results apply more generally.
  Of course, in order to create the thermal bath, there must be another mechanism for the SM particle production. Here, we only consider particle production via the {\it background evolution} in the Dirac equation and thus separate this 
    mechanism from other sources.
  
  Our analysis shows that the main factor in determining the particle production efficiency is the maximal thermal mass, whether it is above or below the condensate-induced effective mass. 
  The result is given by (\ref{density-thermal}) with $M$ being the maximal thermal mass after the Higgs condensate decay.
    For fermions charged under SU(${\cal N}$) and having the Yukawa couplings, the thermal mass is  
  $M^2_f (T)
  = g^2 T^2  {{\cal N}^2-1\over 16 {\cal N}} +  |{ Y}_f |^2 T^2  {N_f \over 16} $, where $N_f$ is the particle multiplicity in the loop.
  Denoting the maximal temperature of the SM thermal bath by $T_{\rm max}$, we find
    \begin{equation}
n_f^{\rm slow} \sim  \alpha^3 g^3  N_c \, T_{\rm max}^3\, {a_e^3\over a^3} \;,
\label{nf-thermal}
\end{equation}
 where we have taken $N=6$ and $M_f(T)= \alpha \times gT/\sqrt{6}$ as the benchmark value for a generic gauge coupling $g$, with $\alpha \sim {\cal O}(1)$ accounting for the different quantum numbers of the SM fermions.
  We observe that this result exhibits a significant degree of universality   and the particle density generated via the Universe expansion on a thermal background can be significantly higher than that
  in  (\ref{nf-step}). This is the case, for example,   when the inflaton decays sufficiently fast generating a large maximal SM temperature \cite{Kolb:1990vq}, 
  $$T_{\rm max} \sim \left(\Gamma_\phi H_e M_{\rm Pl}^2\right)^{1/4} \;,$$ 
  with $\Gamma_\phi$ being the inflaton decay width.
  As in the previous case, the result is essentially independent of the low-energy fermion masses and dictated primarily by the early time particle production, when the temperature is not too far from $T_{\rm max}$.
  We also find that the exact 
  scaling of the temperature with time  is not important, i.e. the results for $T\propto a^{-1}$ and $T\propto a^{-1/2}$
are similar.  
  
  The energy density of the produced particles is bounded by $n_f \times k_*^{\rm th}/a_0$ at the time of the condensate  decay, which again  yields
  $$\rho_f \lesssim M^4_f\;.$$ For $\Gamma_\phi \ll H_e$, this is far below the inflaton energy density, so our approximation is self-consistent.
  Unlike in the previous case, fermion production is almost  democratic and determined primarily by the gauge couplings. Since there is no Yukawa coupling suppression,
  both the number and energy densities can by far exceed those corresponding to the sharp mass decrease case.

\subsubsection{Top-quark production}

In the case of the top-quark, the induced fermion  mass is close to  the Hubble rate and the approximation $M \ll H_e $ breaks down. The main effect of the mass increase  is the change in the inflationary $in$-wavefunctions. As seen from  (\ref{uBinf},\ref{uAinf}),   for $M\gtrsim H_e $, 
the $in$-vector rotates towards  $(1,0)^T$ since at the end of inflation 
$$| u_A^{in} / u_B^{in}| \sim e^{\pi M/H_e} \gg 1 \;,$$
 which makes it
  proportional to the corresponding $out$-vector. Hence, even at small momenta, there is a significant cancellation
  in the Bogolyubov coefficient leading to a smaller average occupation number. The effect is $exponentially$ sensitive to the fermion mass.
  
  For the realistic value of the top Yukawa coupling, such suppression is not too strong and  partially  compensated by the increase in the momentum cutoff compared to that of lighter fermions.
    The result is exponentially sensitive to the precise value of the induced top quark mass, which depends on the poorly constrained $\lambda_h(H_e)$. 
  We choose the benchmark values $M_t = 0.4 H_e$ and $0.7 H_e$, for which we find, assuming 
 the step-function mass term,
 \begin{eqnarray}
  && M_t = 0.4 H_e: ~~|\beta_k|^2 \simeq 0.4  ~~, ~~ k_*^{\rm }\simeq 10^{-4}\, a_e H_e \;, \nonumber\\
  && M_t = 0.7 H_e: ~~|\beta_k|^2 \simeq 0.3  ~~, ~~ k_*^{\rm }\simeq 10^{-3} \, a_e H_e \;.
  \end{eqnarray}
With the smoothed time variation of the effective top-quark mass as motivated by the thermal effects ($M\propto 1/a$), we obtain
   \begin{eqnarray}
  && M_t = 0.4 H_e: ~~|\beta_k|^2 \simeq 0.07  ~~, ~~ k_*^{\rm th}\simeq 0.2\, a_e H_e \;, \nonumber\\
  && M_t = 0.7 H_e: ~~|\beta_k|^2 \simeq 0.01  ~~, ~~ k_*^{\rm th}\simeq 0.6 \, a_e H_e \;.
  \end{eqnarray}
Both the Bogolyubov coefficient and the momentum cutoff change compared to those of lighter fermions. This is expected since  $N \sqrt{M_t/H_e} \gg  1$ and it cannot be used as an expansion parameter.

The resulting top-quark number density with  the step-function $M(\eta)$  dependence is
\begin{equation}
n_t^{\rm fast} \sim 10^{-11} \times N_c \,H_e^3 \, {a_e^3\over a^3} \;,
\end{equation}
 assuming $M_t = 0.7 H_e$ and $N_c$ is the color multiplicity.  Therefore, the top-quark production in this approximation is $suppressed$ compared to that of intermediate mass fermions $b,c,\tau$, but is still more efficient 
 than production of light fermions like $u,d,e$.
 
 For a thermal-like mass profile, the density is much larger,
 \begin{equation}
 n_t^{\rm slow} \sim 10^{-4}\times {N_c }\,H_e^3 \, {a_e^3\over a^3}  \;,
  \end{equation}
  with $M_t = 0.7 H_e$.
This mass value corresponds to a large maximal temperature, around the Hubble scale.
At such a high temperature, all the SM fermion thermal masses are of the same order of magnitude, so the top-quark is not special and the particle abundance is flavor-universal.  
This result shows some suppression of the fermion production when the thermal mass becomes comparable to $H_e$ (cf.\;Eq.\,\ref{nf-thermal}).

\subsubsection{No Higgs condensate}

The main assumption of our work is that a large Higgs condensate forms in the Early Universe. This is not necessarily the case. In the presence of a large effective mass during inflation, the Higgs field rolls to the potential minimum.
This  happens if there is a tangible Higgs-inflaton coupling $\lambda_{\phi h} \varphi^2 h^2$ or a non-minimal Higgs coupling to gravity $\xi h^2 R$, with the appropriate  sign as to generate a positive effective mass term \cite{Lebedev:2012sy}.
In case such a mass term is 
 around or  above the Hubble rate,  the Higgs dynamics are essentially classical and the field is confined to the origin.
 
Formation of the Higgs condensate is essential for particle production in the case of the step-function profile of $M(\eta)$. Thus, $\langle h \rangle \ll H_e$ would eliminate this production mechanism.
On the other hand, if the fermion attains a large thermal mass soon after the end of inflation, its production becomes significant.  The inflationary $in$-wavefunction at $\eta\sim 0$ remains close to $(1/\sqrt{2}, 1/\sqrt{2})^T$, such that 
the results of Sec.\,\ref{thermal-effects} largely apply and the particle number density  has the form (\ref{nf-thermal}).

We note that the existence of the thermal mass does not necessarily imply that the particle itself is part of the thermal bath \cite{Laine:2016hma}.
Indeed, a non-thermal $\nu_R$ can have a substantial thermal mass due to its Yukawa coupling, which represents ``friction'' for the neutrinos propagating in a medium.
 In this case, inflationary particle production via the background evolution can be clearly separated 
from other sources.

Let us summarize our findings so far. 
Inflationary fermion production is sensitive to the Early Universe dynamics such that no precise  prediction can be made. In particular, the production efficiency depends on 
the postinflationary wavefunction. 
The latter is affected by the time-dependent profile 
 of the effective mass function,
which includes thermal/non-equilibrium  effects in addition to the Higgs condensate evolution. In all the cases, however, the energy transfer to the quarks and leptons is very small compared to the inflaton energy.

\subsection{Inflationary production of the right-handed neutrinos due to the Higgs coupling}

Let us now apply our results to production of the right-handed neutrinos.  In what follows, we assume that the $\nu_R$ couplings are small enough such that they do not thermalize
and that $\nu_R$ are not produced directly via inflaton decay.
In this case, there are 3 main sources of neutrinos: gravitational production due to the background evolution, decay of the Higgs condensate and the freeze-in mechanism.
Here we focus on the effects of the Higgs Yukawa coupling ${\cal Y}_\nu$, while the Majorana mass effects will be considered in the next subsection.

\subsubsection{Production of $\nu_R$ due to the background evolution}

In the Early Universe, the right-handed neutrinos attain a tangible mass via their  Higgs couplings ${\cal Y}_\nu$. This can be either due to the formation of the inflationary Higgs condensate or due to the thermal mass of
order ${\cal Y}_\nu T$, which affects neutrino propagation in the thermal background. 
In the latter case, the result (\ref{Y-thermal}) applies with $M \sim {\cal Y}_\nu T/\sqrt{8}$ such that 
\begin{equation}
 Y_{\nu_R}^{\rm grav}  \sim  10^{-2}  \, {\cal Y}_\nu^3\,   {T^3_{\rm max} \over (M_{\rm Pl} H_e)^{3/2}} \;,
 \end{equation}
with $N=6$.
We note that this equation does not assume thermalization of the  right-handed neutrinos\footnote{The $\nu_R$ thermal mass is created by the thermal particles in the loop, that is, the SM leptons and the Higgs field.}   and also applies to a non-thermal environment where 
the effective mass 
 is created by non-equilibrium effects which fade away sufficiently slowly. 

 In the case of the step-function evolution of the neutrino mass, the abundance involves  a higher power of ${\cal Y}_\nu$ and is typically smaller than the above estimate
(cf. Eq.\,\ref{Y-M}).

\subsubsection{Direct neutrino production from the primordial Higgs condensate}

 The Higgs condensate   formed at  the inflationary stage    decays into the SM states, which provides an additional source of quarks, leptons and 
  right-handed neutrinos.  Let us estimate the corresponding $\nu_R$ abundance. 
  
  After the condensate starts oscillating in the $\lambda_h h^4$ potential  at $a\sim a_0$, it can be treated as ``radiation''.
 The energy density of the neutrinos  $\rho_\nu$  is obtained by solving the evolution equations
\begin{eqnarray}
&& \dot \rho_h + 4H \rho_h = -\Gamma_h\,  \rho_h \;,\\
&& \dot \rho_\nu + 4H \rho_\nu = \Gamma_{h_\nu}\,  \rho_h\;,
\end{eqnarray}
 where $\rho_h$ is the Higgs condensate energy density,
 $\Gamma_h$ is its decay width, $\Gamma_{h_\nu}$ is the Higgs condensate decay width into neutrinos, and the factor of 4 indicates the radiation-like scaling of the condensate and $\nu$
 energy density.
 The solution is
 \begin{equation}
 a^4 \rho_\nu =     {\Gamma_{h_\nu} \over \Gamma_h}   \; \left( 1- e^{-\Gamma_h t}     \right)  \: (a^4 \rho_h)_0 \, ,
 \end{equation}
 where $ {\rho_h}_0$ is the initial energy density of the condensate at $a_0$.
The decay happens quite quickly, within an ${\cal O}$(10)-fold increase of the scale factor \cite{Enqvist:2015sua}.     The neutrino share of the initial condensate energy is controlled by the branching ratio  
$${\Gamma_{h_\nu} \over \Gamma_h}  \sim {\cal Y}_\nu^2 \;,$$
which gives the right ballpark estimate for our purposes.
Note that the decay $h \rightarrow \nu_L \nu_R$ is efficient only  for  $$ {\cal Y}_\nu \lesssim \sqrt{\lambda_h} \, ,$$
i.e. when  the neutrinos are lighter than the Higgs.\footnote{For heavier neutrinos, the decay is exponentially suppressed and proceeds via the higher harmonics of the oscillating Higgs field. 
In this case, non-perturbative  fermion production  effects  \cite{Greene:1998nh}     can be significant. }

The initial Higgs condensate energy density is given by the Starobinsky-Yokoyama result \cite{Starobinsky:1994bd},
\begin{equation}
{\rho_h}_0 \sim {3H_e^4 \over 4\pi^2} \;,
\end{equation}
and scales as $a^{-4}$ after the condensate starts oscillating.
The number density of the produced neutrinos can be estimated by taking the average energy of the produced quanta to be given by the effective Higgs mass (see e.g., \cite{Ichikawa:2008ne,Koutroulis:2023fgp}),
$$  E_{\nu_R} \sim m_h^{\rm eff}= \sqrt{3 \lambda_h} \langle h \rangle \;,$$
with  the Higgs condensate  initial value $H_e$ and $\langle h \rangle \propto 1/a$ for $a>a_0$.
At the time of the condensate decay $a=a_d$\footnote{In our previous  fermion production analysis, we have loosely associated  $a_0$ with the condensate oscillation and decay, whereas  here we separate $a_0$ and $a_d$.  }, we get
\begin{equation}
n_\nu (a_d) \sim \left({a_0\over a_d}\right)^3 \, {\Gamma_{h \nu} \over \Gamma_h} \, H_e^3 \;,
\end{equation}
for $\lambda_h \sim 10^{-2}$.
After that, the total number of $\nu_R$ is conserved, if we neglect other sources of $\nu_R$-production.

The $\nu_R$ abundance is given by
$
Y_{\nu_R} = {n_{\nu_R} \over s_{\rm SM}} \;,
$
which remains  constant after the Higgs condensate has decayed and reheating has completed. The scaling of the numerator and the denominator with $a$ is the same, hence $Y_{\nu_R}$ 
can be computed at any convenient point, e.g. $a=a_d$. 
The result is
\begin{equation}
Y_{\nu_R}^{\rm higgs} \sim 10^{-1} N^3 \, {\cal Y}_\nu^2 \, \left({H_e\over M_{\rm Pl}}\right)^{3/2} \;,
\end{equation}
where the superscript refers to the direct $\nu_R$ production from Higgs decay.
As long as $T_{\rm max} \lesssim H_e$, this is much  larger than the neutrino abundance produced gravitationally $Y_{\nu_R}^{\rm grav}$, with the inclusion of the thermal-like  mass. The latter is suppressed by an additional power of ${\cal Y}_{\nu}$ and 
$$    { Y_{\nu_R}^{\rm grav}  \over Y_{\nu_R}^{\rm higgs} }  < {\cal Y}_{\nu} \ll 1 \;.$$ 
Hence, in this framework, gravitational fermion production gives only a subleading result. 
 
 \subsubsection{Freeze-in $\nu_R$ production}
 
Both contributions are dwarfed by the usual freeze-in production \cite{Dodelson:1993je}, in which case the abundance scales as \cite{Hall:2009bx}
 \begin{equation}
Y_{\nu_R}^{\rm FI} \sim   {\cal Y}_\nu^2 \,  {M_{\rm Pl} \over m_h} \;,
\label{FI}
\end{equation}
for $m_h \gg M_{\nu_R}$, where $M_{\nu_R}$ is the low energy neutrino mass scale.
 The production  mode is $h \rightarrow \nu_L \nu_R$,
which becomes most efficient 
  at $T\sim m_h$. For   $m_h \lesssim  M_{\nu_R}$, the neutrino production
proceeds via $hh \rightarrow \nu_R \nu_R$, which is most efficient at $T \sim M_{\nu_R}$ such that the scaling (\ref{FI}) holds up to the replacement $m_h \rightarrow M_{\nu_R}$.
 
 For a very heavy $\nu_R$,   $M_{\nu_R} \gg m_h$,    the freeze-in production involves a higher power of ${\cal Y}_\nu$ and it could also be Boltzmann-suppressed \cite{Cosme:2023xpa}. In this case, the Higgs condensate decay can  be a competitive  neutrino source.

Our conclusion is that gravitational $\nu_R$ production is superseded by that from the Higgs condensate decay. The latter, in turn, is  swamped by the standard freeze-in production, unless the right-handed neutrinos are very heavy.
Therefore, we find the following hierarchy, 
$$   Y_{\nu_R}^{\rm grav}  \ll Y_{\nu_R}^{\rm higgs}  \ll Y_{\nu_R}^{\rm FI} \;.$$

\subsection{Inflationary  right-handed neutrino production due to the singlet scalar coupling}

The inflationary  right-handed neutrino production due to the Higgs Yukawa couplings cannot be the leading source of $\nu_R$ in typical cosmological settings. However, this conclusion does not apply to the $\nu_R$ coupling to 
a singlet scalar $s$. Unlike the Higgs field, the new scalar can have a very small  self-coupling, which 
suppresses the $\nu_R$ production from the condensate decay and also 
precludes thermalization. In this case, gravitational production can be the leading source of the right-handed neutrinos.

The scalar condensate can be long-lived, $N\gg 1$, which invalidates our expansion in $N \, \sqrt{M/H_e}$. Let us consider the large $N$ case more closely,
assuming a non-thermal system and a step-function fermion mass variation.
  
 \begin{figure}[h!]
    \centering
    \includegraphics[width=0.6\textwidth]{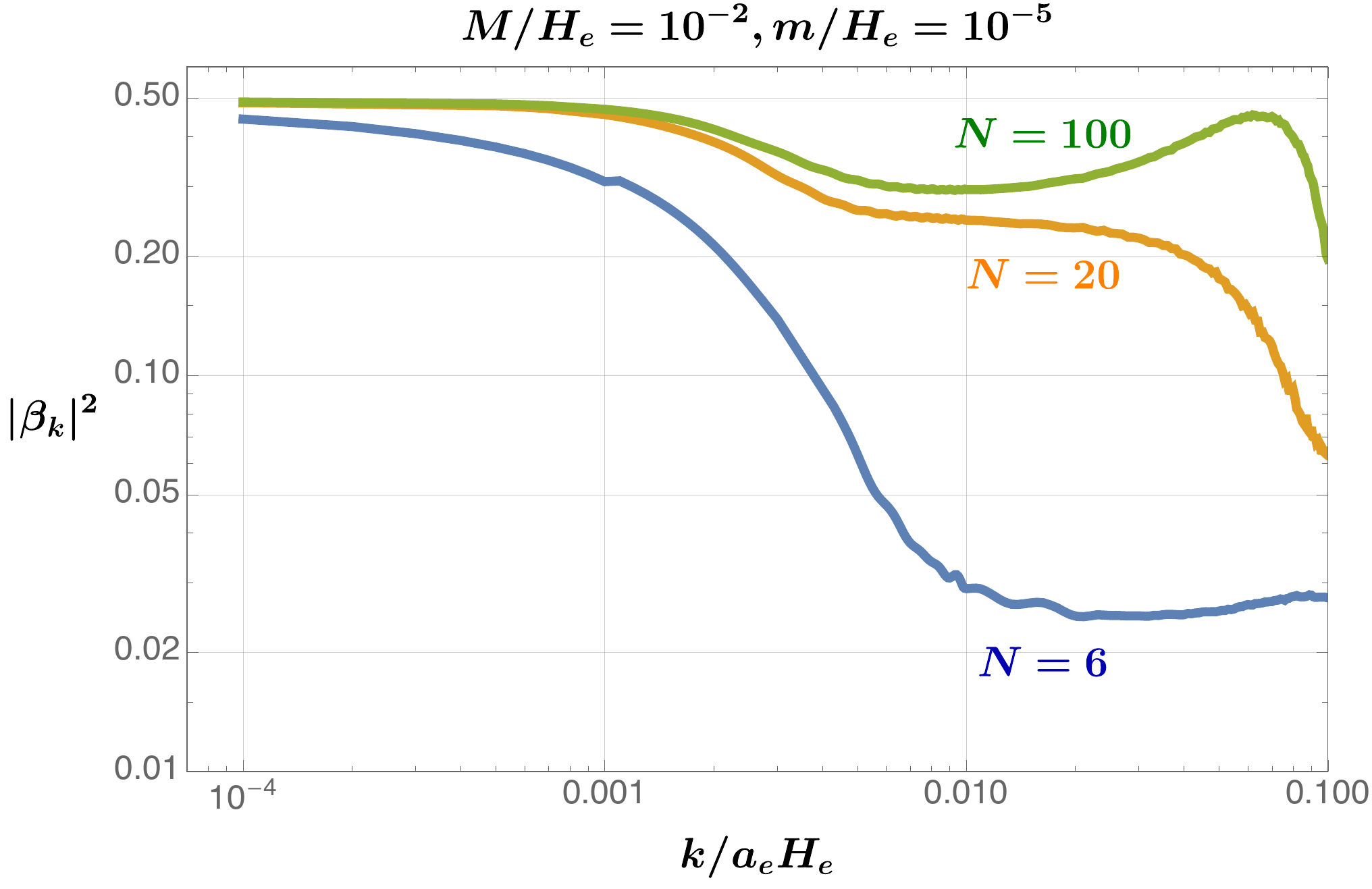}
    \caption{Bogolyubov coefficient squared for $N=6,20,100$ and step-function mass variation. Radiation domination after inflation is assumed
    and $m= 10^{-5} H_e\,,\, M= 10^{-2} H_e\,.$ At large $N$, the curves approach the single mass $|\beta_k|^2$ with the momentum cutoff $M^{1/2} H_e^{1/2} a_e$.}
    \label{large-N}
\end{figure}

\subsubsection{Large $N$ limit for fermion production }

Suppose 
\begin{equation}
N \; \sqrt{M\over H_e} \gtrsim 1\;,
\end{equation}
which means that the fermion mass is above the Hubble rate at the time of the condensate decay.
Clearly, our expansion in $N \, \sqrt{M/H_e}$ is no longer valid. On the other hand, at large $N$, 
the fermion mass remains constant for a long time and 
we approach a single mass case, which is well understood. 
In particular, 
\begin{equation}
|\beta_k|^2 \simeq 1/2 ~~{\rm for }~~ k\lesssim M^{1/2} H_e^{1/2} a_e \;,
\label{single-m}
\end{equation}
and zero above this cutoff.
One 
expects the transition to the single-mass case to take place at  $N^2 \, M/H_e \sim {\cal O}(1)$ and this is indeed what we observe numerically.  
Fig.\,\ref{large-N} displays $|\beta_k|^2$ for three values of $N$, which correspond to $N \, \sqrt{M/H_e} =0.6, \, 2, \, 10$. 
We observe that the curve for $N \, \sqrt{M/H_e} = 10$ closely resembles that for  the single mass $(M)$ case, with the same momentum cutoff 
$M^{1/2} H_e^{1/2} a_e$.

Hence, as long as $N \, \sqrt{M/H_e} \gg 1$,  we can approximate our system by the single mass case (\ref{single-m}).

\subsubsection{Upper bound on the abundance of gravitationally produced fermions}
\label{upper-bound}

In the large $N$ limit corresponding to a long-lived scalar condensate $\langle s \rangle$, we recover the single-mass abundance result,
\begin{equation} 
   Y_0\simeq 5\times 10^{-3} \; \left({M\over M_{\rm Pl}}\right)^{3/2} \;,
    \end{equation}
where $M$ is the fermion mass generated by the Yukawa coupling to the scalar $s$, $M=  {\cal Y}^s \,  \langle s \rangle$.
This result applies to $M \ll H_e$, while for heavier fermions, $M \gtrsim H_e$, the abundance is suppressed.
Using the PLANCK/BICEP bound $H_e \lesssim 10^{13}\,$GeV \cite{Planck:2018jri,BICEP:2021xfz}, we thus obtain the upper bound on the fermion abundance,
\begin{equation} 
   Y_{\rm max} \lesssim 4\times 10^{-11}\;.
    \end{equation}
The dark matter abundance is $4.4\times10^{-10}\,  {\rm GeV} / M_{\rm DM}$, hence
only fermions heavier than about 10 GeV can play the role of dark matter,
\begin{equation} 
 M_{\rm DM} \gtrsim 10\; {\rm GeV} \;,
    \end{equation}
assuming that  they are  produced via inflation. There are further constraints on such fermions due to the isocurvature perturbations \cite{Herring:2020cah}. 

Note that the late time dynamics of the condensate does not play a role as long as   $N  \sqrt{M\over H_e} \gg 1\;.$
The above bound also applies to the matter dominated Universe since the abundance  in this case is further suppressed by $(T_R/M_{\rm Pl})^\alpha$,
with some positive $\alpha$, as we show in the subsequent section. 

Throughout this work, including the above estimate, we have used the ``smoothed'' scale factor function $a(\eta)$. In reality, $a(\eta)$ contains small oscillations 
induced by oscillations of the inflaton field after inflation \cite{Ema:2015dka}. 
Their effect is encapsulated by the effective Planck-suppressed operator that couples the inflaton $\phi$ to the fermion,
\begin{equation}
{\cal C}\, {M \over M_{\rm Pl}^2 }\, \phi^2 \bar \Psi \Psi\;,
\end{equation}
 where ${\cal C} \sim 10^{-1}$. 
 One may estimate particle production via this operator by assuming that $M$ is constant during the inflaton oscillations  and $M \ll m_\phi$. 
The corresponding perturbative production rate  has been computed in \cite{Koutroulis:2023fgp}. Taking the inflaton field value at the beginning of oscillations $\phi_0 \sim M_{\rm Pl}$ and $M\sim H_e \sim 10^{13}\,$GeV,
 one finds that this mechanism is somewhat less efficient and the       corresponding bound on $m_{\rm DM}$ is   in the ballpark of 100 GeV.
  This shows that fermion production by an oscillating inflaton can be competitive, depending on further details. Note that this effect is due to {\it classical} gravity.

The above dim-5 operator is also produced by quantum gravity effects \cite{Koutroulis:2023fgp}, although with an unknown coefficient. After inflation, one may use the effective field theory expansion to account for gravity-induced operators  \cite{Lebedev:2022ljz}  and compute their
contribution to particle production. The result is  that such operators  can be very efficient and, thus, the above bound only applies to classical gravity.

\subsubsection{Feebly interacting scalar extension}

Let us now consider an example of a minimalistic model, where a singlet fermion is produced predominantly by inflation.
We identify 
 the fermion with a right-handed neutrino and 
 add one more degree of freedom \cite{Kusenko:2006rh}: a light real scalar $s$ with the potential
\begin{equation}
V(s)= {1\over 2} \mu_s^2 \, s^2 +{1\over 4} \lambda_s \, s^4 \;,
\end{equation}
and the coupling to $\nu_R$,
\begin{equation}
\Delta {\cal L}={1\over 2}\, {\cal Y}^s \, s \,\nu_R \nu_R + {\rm h.c.}
\end{equation}
The effective right-handed neutrino mass takes on different values at early and late times, as determined by $\langle s \rangle$, while the Dirac neutrino mass is negligible. In addition, $\nu_R$ can also have a rigid Majorana mass term contribution, although this plays no significant role.
 We take the scalar self-coupling to be small, $\lambda_s \ll 1$, and its coupling to the Standard Model, e.g. the Higgs field \cite{Lebedev:2021xey}, to be feeble. The self-coupling cannot be arbitrarily small since it is generated at one loop via the fermion loop, hence
to avoid fine-tuning, we require 
\begin{equation}
\lambda_s > {1\over 8\pi^2 } \left|  {\cal Y}^s \right|^4 \;.
\label{stability}
\end{equation}

A light scalar field  $s$ experiences large quantum fluctuations during inflation. The asymptotic value,
\begin{equation}
\langle s^2 \rangle \rightarrow  0.1 {H^2 \over \sqrt{\lambda_s}} \;,
\end{equation}
is reached within the  characteristic relaxation  time $(\sqrt{\lambda_s } H)^{-1}$ \cite{Starobinsky:1994bd}.
When $\lambda_s$ is small and the duration of inflation is finite, the field does not have enough time to relax to the asymptotic value. In this case, the field value at the end of inflation 
is determined primarily by the pre-inflationary initial conditions \cite{Feiteira:2025rpe}. The result can  then be parametrized in terms of the condensate value at the end of inflation $s_e \equiv \sqrt{\langle s^2 (a_e) \rangle}$.

After inflation, the average   field $s$ satisfies
$$  \ddot s + 3H \dot s + V^\prime_s =0\;.       $$
 As long as the effective mass is smaller than $H$, the last term can be neglected such that  $\dot s \simeq 0$ is a solution to the EOM with the $\dot s=0$
 initial condition. Hence, the condensate size remains to be given by $s_e$ for some time after inflation.
 
 The condensate starts oscillating in a quartic potential when  the potential term becomes important, i.e. the Hubble rate reduces to 
 \begin{equation}
 H_0 \sim m_s^{\rm eff} =  \sqrt{3\lambda_s}\, s_e
 \label{osc-s}
 \end{equation}
 Its contribution to the total energy density is  assumed to be subdominant, $V(s) \ll 3 H^2 M_{\rm Pl}^2 $. Applying this condition at the start of the oscillation period, we find
  \begin{equation}
   s_e \ll M_{\rm Pl} \;,
 \end{equation}
 required for consistency of our analysis. The lower bound $s_e \gtrsim H_e$ is imposed by the scalar fluctuation analysis \cite{Feiteira:2025rpe}.
 The condensate oscillation leads to particle production, i.e. conversion of the zero mode into $s$-quanta \cite{Greene:1997fu,Khlebnikov:1996mc}. If the effective neutrino mass is much larger than the effective scalar mass,
 \begin{equation}
   {\cal Y}^s \gg \  \sqrt{\lambda_s} \;,
   \label{ys-lambda}
 \end{equation}
the $\nu_R$ production from the condensate is highly suppressed kinematically. This is consistent with the radiative stability condition (\ref{stability}) for $\lambda_s\lesssim 1$. 
Hence, the neutrino production through the condensate decay can be neglected. 
Similarly, $s$ decay into neutrinos at late times  is forbidden 
  as long as the bare mass of $\nu_R$ is larger than that of $s$.

For small enough couplings, $s$ and $\nu_R$ do not thermalize, nor is there the freeze-in contribution from the SM thermal bath. Hence, the main contribution to the $\nu_R$ abundance 
comes from gravitational fermion production. 
 It is efficient for $M_\nu={\cal Y}^s s_e  \lesssim H_e$ and $N\gg 1$. Specifically,  the condensate starts oscillating at $H_0 \simeq H_e /N^2$ which, combined with (\ref{osc-s}) 
 and $s_e = M_\nu/{\cal Y}^s$,  
 requires
 \begin{equation}
 {M_\nu \over H_e} N^2 \sim   {{\cal Y}^s \over \sqrt{\lambda_s}} \gg 1 \;,
 \label{N}
 \end{equation}
   by virtue of (\ref{ys-lambda}). This ensures that the condensate is indeed long-lived and the single-mass approximation is adequate.
   
   Ignoring the difference between the Dirac and Majorana fermions, which is within  the error bars of our calculation, the $\nu_R$ abundance is given by 
   \begin{equation} 
   Y_\nu \sim  5\times 10^{-3} \;    \,\left({{\cal Y}^s s_e\over M_{\rm Pl}}\right)^{3/2} \;,
    \end{equation}
   which matches the dark matter abundance for 
   \begin{equation} 
   m_\nu \sim 10 \; {\rm GeV} \times \left(   {10^{13} \, {\rm GeV}   \over   {\cal Y}^s s_e  }   \right)^{3/2} \;,
       \end{equation}
  with   
    ${\cal Y}^s s_e\lesssim  H_e \lesssim 10^{13}\,$GeV. The low energy right-handed neutrino mass $m_\nu$ is determined by  the VEV of the scalar singlet, possibly with the contribution of the 
    rigid mass term. In order to be viable dark matter candidates, these neutrinos must also satisfy the isocurvature constraints, which will be studied elsewhere. Furthermore, ${\cal Y}^s$ must be sufficiently small
    to avoid thermalization of the system\footnote{Thermalization of this system was considered in 
    \cite{DeRomeri:2020wng,Lebedev:2023uzp}, although the results do not apply directly to the case at hand due to a highly non-thermal initial state and $m_\nu \gg m_s$.}, while $ s_e$ is constrained to be between $H_e$ and  $M_{\rm Pl}$.

   The quanta of $s$ do not contribute to the dark matter energy density if they decay into light SM particles after the condensate break-up. This can happen due to higher dimension operators like ${s} F_{\mu\nu} F^{\mu\nu}, {\rm  \;etc}.$ or
   a tiny mixing with the Higgs field. 
    
    This example shows that inflationary fermion production can be important and even the leading mechanism for production of very weakly interacting particles.

  \section{Inflation followed by  the matter dominated era}
  \label{matter}
  
 The postinflationary dynamics can take place in a matter-dominated background instead of the radiation-dominated one. This is the case for a $\phi^2$ $local$ inflaton potential before reheating, which may happen at a much later stage. 
  In the matter dominated epoch ($\eta>0$), 
    \begin{equation}
 H=H_e \left( {a_e\over a}\right)^{3/2} ~~,~~ a= {1\over 4} a_e^3 H_e^2 \left( \eta + {2\over a_e H_e}   \right)^2 \;,
 \end{equation}
 which connects smoothly to the inflationary regime at $\eta=0$.
We assume  that the Universe has the matter-like equation of state, e.g. it is dominated by the non-relativistic inflaton, for an extended period such that 
the boundary conditions in the $out$-region are  imposed during this  era. Eventually, radiation takes over the energy balance, albeit 
without affecting the particle number.

  Since we are only interested in the $out$-region,  in what follows, $u_{A,B}$ will refer to  $u_{A,B}^{\rm out}$. The $in$ wavefunctions remain intact.

   \subsection{Constant mass}

   At $\eta \gg 1/(a_e H_e)$, the EOM for $u_A$ reads
 \begin{equation}
 u_A^{\prime \prime} + \left(     {i\over 2}  m H_e^2 a_e^3\,\eta   +  {1\over 16} m^2 H_e^4 a_e^6\, \eta^4 +k^2   \right)\, u_A =0 \;,
 \end{equation}
 while the EOM for $u_B$ is obtained by replacing  $m \rightarrow -m$.
In the $k^2 \rightarrow 0$ limit, the solution can be expressed via hypergeometric functions \cite{diff-eq}, while in the general case, finding an analytical solution is challenging. 
   
   The oscillation frequency at late times is $\omega \simeq a m = {1\over 4} m H_e^2 a_e^3 \, \eta^2$ such that the boundary condition at $\eta \rightarrow \infty$ becomes
 \begin{equation}
 \left(
 \begin{matrix}
  u_A \\
  u_B
 \end{matrix}
 \right)^{out}  \simeq 
 \left(
 \begin{matrix}
       1 \\
  {2k\over  m H_e^2 a_e^3 \, \eta^2} 
 \end{matrix}
 \right) \times    e^{ -{i\over 12 } m H_e^2 a_e^3 \, \eta^3}    \;,
 \end{equation}
 to linear order in $k$.
Using the  inflationary $in$ states as before, one  computes $\beta_k^2$   at $\eta\sim 0$ finding that the effective momentum  cut-off for particle production is
  \begin{equation}
k_*(m) = m^{1/3} H_e^{2/3} a_e \;.
\label{k-matter}
 \end{equation}
 This corresponds to the momentum at which the three contributions to the frequency become comparable, $m H_e^2 a_e^3\,\eta \sim m^2 H_e^4 a_e^6\, \eta^4 \sim k^2 \;.$
 It can also be written as $k_* = a_m m$, where $a_m$ is the scale factor at which $H(a_m)=m$, implying that the produced particles are non- or semi-relativistic at $a=a_m$.  
 Below the cutoff, $|\beta_k|^2$ is approximately $1/2$, while above it, the Bogolyubov coefficient  goes to zero.

    \subsection{Step-function mass term}
   
   Let us now take the mass term of the form $M\, \theta(\eta_0 -\eta) + m\, \theta(\eta -\eta_0) $ with $M\gg m$. 
   Here $\eta_0$ corresponds to the Higgs condensate decay  time and 
   \begin{equation}
   a_0/a_e =N \;,
   \end{equation}
    as before.
   
   The wave-function at late times is described above, while at  $\eta < \eta_0$ it
   satisfies 
    \begin{equation}
 u_A^{\prime \prime} + \left(     {i\over 2}  M H_e^2 a_e^3\,\eta   +  {1\over 16} M^2 H_e^4 a_e^6\, \eta^4 +k^2   \right)\, u_A =0 \;,
 \end{equation}
with proper boundary conditions at $\eta_0$,
\begin{equation}
  u'_A\Bigl\vert_- =  u'_A\Bigr\vert_+    -ia_0(M-m)\,u_A (\eta_0) \;,
   \label{bound-cond}
  \end{equation}
The EOM for $u_B$ and the corresponding boundary conditions  are obtained by replacing  $M,m \rightarrow -M,m$.
   
The main features of the solution can be understood analytically.   
At early times, the $\eta^4$ term in the EOM  is dominated by the linear term in $\eta$  and can be neglected  as long as $M/H_e \times N^{3/2} \ll 1$.   
 Then, the wavefunction satisfies a simple equation at $\eta < \eta_0$,
    \begin{equation}
 u_A^{\prime \prime} +   \left(  {i\over 2}  M H_e^2 a_e^3\,\eta +k^2 \right) \, u_A =0 \;,
 \end{equation}
   whose solution is a linear combination of the Airy functions. 
   The corresponding equation for $u_B$ is obtained by the replacement $M \rightarrow -M$.
   At small momenta, $k^2 \ll \sqrt{N}\, MH_e a_e^2$,
    \begin{eqnarray}
 &&u_A (\eta) = a_1 \, {\rm Ai} \left(k_*(M) \,\eta /(2i)^{1/3}\right) + a_2 \, {\rm Bi} \left(k_*(M)\, \eta /(2i)^{1/3}\right) \;,\\
 &&  u_B (\eta) = b_1 \, {\rm Ai} \left(k_*(M)\, \eta /(-2i)^{1/3}\right) + b_2 \, {\rm Bi} \left(k_*(M)\, \eta /(-2i)^{1/3}\right) \;,
 \end{eqnarray}
 where $k_*(M)$ is given by (\ref{k-matter}) with the replacement $m\rightarrow M$, and $a_i , b_i$ are constant coefficients.
   Since $\eta_0 \simeq \sqrt{N}\, \eta_e$ and $\eta_e \simeq 2/(a_e H_e)$, 
   the argument of the Airy function is a small number,  $k_* \eta \sim  \sqrt{N} \; (M/H_e)^{1/3}  \ll 1$.
   Hence, one can use the expansion 
   \begin{eqnarray}
 && {\rm Ai} (x) \simeq {1\over 3^{2/3} \Gamma (2/3) } \left( 1+ x^3/6 \right) -  {x\over 3^{1/3} \Gamma (1/3) } \;,\\
 &&   {\rm Bi} (x) \simeq {1\over 3^{1/6} \Gamma (2/3) } \left( 1+ x^3/6 \right) +  { 3^{1/6} \,x\over  \Gamma (1/3) }      \;.
 \end{eqnarray}
   The coefficients $a_i$ and $b_i$ are fixed by the boundary conditions at $\eta_0$. 
   
   For momenta below the small-mass cutoff $k_*(m)$, nothing changes compared to the single mass case. We are interested in the higher momenta modes $k\gg k_*(m)$, 
   which bring in new effects. In this case, $u_A$ and $u_B$ are essentially the same at early times, in particular at $\eta_0 +\epsilon$. This is because the corresponding
   modes should give $\beta_k \simeq 0$ in the single mass case, which implies 
   $$u_{A,B} \left(\eta\gtrsim \eta_0\right) \simeq {1\over \sqrt{2}} \, e^{-i k \eta} \, ,$$ 
   with $\omega\simeq k$, neglecting the terms proportional to $m$.
   This, together with the derivative jumps (\ref{bound-cond}), determines the boundary condition at $\eta_0$ and fixes $a_i , b_i$.

   The Bogolyubov coefficient can be estimated by approximating $u_{A,B} (\eta_e)$ with $u_{A,B} (0)$. We find that the linear and cubic terms in $\eta_0$ in the expansion of the Airy 
   functions give contributions of the same order in $M/H_e$. The universal phase proportional to $k\eta_0$ cancels in $|\beta_k|$ and we get
   \begin{equation}
   |\beta_k| \simeq {1\over \sqrt{2}} \left|u_A^{\rm out}(0) - u_B^{\rm out} (0) \right| \sim {M\over H_e} \; N^{3/2}\;,
   \end{equation} 
  in the momentum range $\sqrt{ \sqrt{N} M H_e a_e^2}  \gg k\gg k_* (m)$. Here we have neglected the logarithmic momentum dependence coming from the $in$ states. 
  At higher momenta, the phase attained by $u_{A,B}$ during the evolution from $\eta_0$ to $\eta_e$     becomes
  universal and the Bogolyubov coefficient gets suppressed by powers of $M/k$. Thus, the momentum cutoff for the above expression is 
   \begin{equation}
   \tilde k_* \simeq N^{1/4 } \times  M^{1/2} H_e^{1/2} a_e\;.
   \end{equation} 
   
   The resulting fermion abundance is obtained by noting that the particle number is conserved at late times and  given by $n a_R^3$, where $a_R$ is the scale factor at reheating.   Since
   $a_R^3 = a_e^3 \, H_e^2/H_R^2$, one finds at $m\ll M$,
   \begin{equation}
   Y \sim 10^{-2} \times N^{15/4} \, {M^{7/2} T_R \over H_e^{5/2} M_{\rm Pl}^2} \;,
   \end{equation}
   where $T_R$ is the reheating temperature.
   For the matter dominated case,  we may take 
   $N\sim 10 $ as the benchmark number corresponding to $N=6$ of the radiation dominated case.  The result is suppressed by $T_R/\sqrt{H_e M_{\rm Pl}} \ll 1$ 
   compared to $Y$ generated during the radiation dominated era.   
    
    \subsubsection{Universal form}
    
    It is interesting to note that  the radiation and matter domination results can be put in a {\it universal form} which involves the Hubble rate at the time of the Higgs condensate decay $H_0$ instead of $N$.
    The Bogolyubov coefficient  in both  cases has the form  $M/H_e \times H_e/H_{0}\,$. Hence, we get the universal result
\begin{equation}
|\beta_k| \sim {M\over H_{0}} \;,
\end{equation}   
together with the universal cutoff which can be put in the form
\begin{equation}
\tilde k_*/a_0 \sim  \sqrt{M H_0}\;.
\end{equation}   
The abundance then reads
\begin{equation}
Y \sim \left( 10^{-3}-10^{-2}\right) \times {M^{7/2} \over  H_0^2 \,M_{\rm Pl}^{3/2}} \times  \left(  {T_R \over \sqrt{H_0 M_{\rm Pl}} } \right)^\gamma \;,
\label{univ-Y}
\end{equation}
where $\gamma$ is 0 and 1 for the radiation and matter dominated cases, respectively, and
the prefactor uncertainty represents the typical ``error-bars'' expected in our calculation.
  Note that $H_0$ is fixed by the effective scalar mass at the end of inflation, $m_{\rm eff} \sim \sqrt{3\lambda_h}\langle  h\rangle$. 
This expression clearly shows the suppression of $Y$ in the matter dominated case: $ {T_R \over \sqrt{H_0 M_{\rm Pl}} } < 1$ since reheating occurs after the condensate decay by definition.

Similarly, the single mass result ($N \gg 1$) can be put in the universal form,
   \begin{equation} 
   Y_0\simeq 10^{-2}\;  \left({M\over M_{\rm Pl}}\right)^{3/2} \times \left(  {T_R \over \sqrt{M M_{\rm Pl}} } \right)^\gamma\;,
    \end{equation}
    where $\gamma=0$ or 1, as in  (\ref{univ-Y}).
  Again,  ${T_R \over \sqrt{M M_{\rm Pl}} } <1$ by assumption of a long matter-dominated period. 

 The $T_R$ suppression of the abundance  is a common feature of particle production in the matter dominated epoch. Indeed, if particle production is dominated by early times,
 $$Y \simeq { {\rm const}\over T_R^3 a_R^3} \propto H_R^{1/2} \propto T_R\;.$$ 
  On the other hand, in the radiation domination case, the $T_R$ dependence cancels out.
This conclusion also applies to smooth $M(\eta) $ functions, e.g. the thermal mass, hence the particle abundance in the matter dominated case is suppressed compared to that
in the radiation dominated scenario.     

Given the above  universal suppression factor, we conclude that particle production is more efficient in the radiation-dominated background and the particle abundance produced in the  matter-dominated     
case can be made very small by reducing the reheating temperature, which is only bounded by 4 MeV from below \cite{Hannestad:2004px}.

   \section{Conclusion}

   We have studied inflationary production of fermions in realistic cosmological settings. The production efficiency is determined by the fermion mass, which is time-dependent and can be very large in the Early Universe.
   For example, the SM fermion mass is controlled by the Higgs field value. In the absence of significant couplings to the inflaton, scalar fields experience large fluctuations during inflation, which drive the average field value to the Hubble scale and above. Thus, quarks and leptons
   were many orders of magnitude heavier during inflation, compared to their current masses. This dramatically increases efficiency of their gravitational production.

   Using the Bogolyubov coefficient approach, we obtain general results for gravitational production of fermions with 
   sharp and slow mass variations, which we model by a step-function and a thermal mass time dependence.     
      For the sharp mass decrease, the resulting particle abundance scales as 
      $$Y \propto  {M^{7/2} \over M_{\rm Pl}^{3/2} H_e^2}\, ,$$
   while the slow mass decrease results in $$Y \propto  {M^{3} \over (M_{\rm Pl} H_e)^{3/2} }\;, $$ where $M$ the effective fermion mass shortly
   after inflation and $H_e \gg M$ is the Hubble rate at the end of inflation. These results assume radiation domination after inflation, while
   in the case of matter domination, the abundance is suppressed by an additional factor depending on the reheating temperature.

   Applying our  results to the SM fermions, we find that the production efficiency grows by many orders of magnitude compared to that based on the constant low energy masses.
   Nevertheless, the 
   energy density of the produced particles remains small and does not affect the standard approach to reheating. However, the production mechanism can be relevant to the scenarios where the inflaton decays entirely into the dark sector states
   by creating an irreducible SM background.
   These considerations also apply to  production of the right-handed neutrinos with the Higgs Yukawa couplings. Such neutrinos  are produced by inflation, via the Higgs condensate decay and through the freeze-in mechanism. We find that gravitational particle production is (at best) subleading 
   in this case.

  Inflationary fermion production can also be the leading particle source. For example,  
   if the the right-handed neutrinos couple to a light singlet scalar, it  induces a large Majorana neutrino mass in the Early Universe. For small enough couplings, the resulting gravitational particle production  
   dwarfs other sources and can account for the dark matter abundance. This scenario is subject to the isocurvature constraints to be studied in a subsequent publication.
   
   We find the following general result: inflationary expansion can be responsible for the fermionic dark matter abundance only if the (low energy) fermion mass is bounded by
    $$  m_{\rm DM} \gtrsim 10\; {\rm GeV} \;.    $$
   This bound assumes a smooth transition from inflation to 
    radiation/matter domination and is based on the  Hubble rate constraint $H_e \lesssim 10^{13}\,$GeV. 
   Therefore, {\it classical} gravity cannot produce a sufficient number of keV-scale sterile neutrinos.
   In contrast, quantum gravity-induced operators can account for cold, keV sterile neutrino dark matter free of the isocurvature constraints  \cite{Koutroulis:2023fgp}.
   \\ \ \\
   {\bf Acknowledgements.} OL acknowledges support from the Magnus Ehrnrooth  foundation. The work of FK is supported in part by the National Natural Science Foundation of China under grant No.
12342502.


\begin{thebibliography}{99}
   
     
\bibitem{Mukhanov:2005sc}
V.~Mukhanov,
``Physical Foundations of Cosmology,'' Cambridge University Press, 2005;
doi:10.1017/CBO9780511790553.

\bibitem{LZ:2024zvo}
J.~Aalbers \textit{et al.} [LZ],
Phys. Rev. Lett. \textbf{135}, no.1, 011802 (2025).
   
   
   
   
\bibitem{Parker:1969au}
L.~Parker,
Phys. Rev. \textbf{183}, 1057-1068 (1969);
A.~A.~Grib and S.~G.~Mamaev,
Yad. Fiz. \textbf{10}, 1276-1281 (1969);
Y.~B.~Zeldovich and A.~A.~Starobinsky,
Zh. Eksp. Teor. Fiz. \textbf{61}, 2161-2175 (1971)

\bibitem{Parker:1971pt}
L.~Parker,
Phys. Rev. D \textbf{3}, 346-356 (1971)
[erratum: Phys. Rev. D \textbf{3}, 2546-2546 (1971)].

\bibitem{Grib:1976pw}
S.~G.~Mamaev, V.~M.~Mostepanenko and A.~A.~Starobinsky,
Zh. Eksp. Teor. Fiz. \textbf{70}, 1577-1591 (1976);
A.~A.~Grib, S.~G.~Mamaev and V.~M.~Mostepanenko,
Gen. Rel. Grav. \textbf{7}, 535-547 (1976).

      
 

   
   
   
      \bibitem{Lebedev:2022cic}
O.~Lebedev,
JCAP \textbf{02}, 032 (2023).

   
   
\bibitem{Starobinsky:1980te}
A.~A.~Starobinsky,
Phys. Lett. B \textbf{91} (1980) 99-102.


\bibitem{Guth:1980zm}
A.~H.~Guth,
Phys. Rev. D \textbf{23} (1981) 347-356.

\bibitem{Linde:1981mu}
A.~D.~Linde,
Phys. Lett. B \textbf{108} (1982) 389-393; Phys. Lett. B \textbf{129} (1983), 177-181.


   
   
\bibitem{Bogolyubov:1958se}
N.~N.~Bogolyubov,
Sov. Phys. JETP \textbf{7}, 41-46 (1958)
JINR-R-94.
   
   
\bibitem{Chung:2011ck}
D.~J.~H.~Chung, L.~L.~Everett, H.~Yoo and P.~Zhou,
Phys. Lett. B \textbf{712}, 147-154 (2012).

\bibitem{Koutroulis:2023fgp}
F.~Koutroulis, O.~Lebedev and S.~Pokorski,
JHEP \textbf{04}, 027 (2024).

   
\bibitem{Starobinsky:1986fx}
A.~A.~Starobinsky,
Lect. Notes Phys. \textbf{246}, 107-126 (1986).

   
   
   
\bibitem{Starobinsky:1994bd}
A.~A.~Starobinsky and J.~Yokoyama,
Phys. Rev. D \textbf{50}, 6357-6368 (1994).
   
   
   
\bibitem{Boyarsky:2009ix}
A.~Boyarsky, O.~Ruchayskiy and M.~Shaposhnikov,
Ann. Rev. Nucl. Part. Sci. \textbf{59}, 191-214 (2009).
   
   
\bibitem{Boyarsky:2018tvu}
A.~Boyarsky, M.~Drewes, T.~Lasserre, S.~Mertens and O.~Ruchayskiy,
Prog. Part. Nucl. Phys. \textbf{104}, 1-45 (2019).
   
\bibitem{Ford:2021syk}
L.~H.~Ford,
Rept. Prog. Phys. \textbf{84}, no.11, 116901 (2021).



\bibitem{Kolb:2023ydq}
E.~W.~Kolb and A.~J.~Long,
Rev. Mod. Phys. \textbf{96}, no.4, 045005 (2024).

  
\bibitem{Bunch:1978yq}
T.~S.~Bunch and P.~C.~W.~Davies,
Proc. Roy. Soc. Lond. A \textbf{360}, 117-134 (1978).



   
   
   
   
   
\bibitem{Herring:2020cah}
N.~Herring and D.~Boyanovsky,
Phys. Rev. D \textbf{101}, no.12, 123522 (2020).
   
\bibitem{Klaric:2022qly}
J.~Klaric, A.~Shkerin and G.~Vacalis,
JCAP \textbf{02}, 034 (2023).
   
\bibitem{Barman:2022qgt}
B.~Barman, S.~Cl{\'e}ry, R.~T.~Co, Y.~Mambrini and K.~A.~Olive,
JHEP \textbf{12}, 072 (2022).
   
   
   
\bibitem{Feiteira:2025rpe}
D.~Feiteira and O.~Lebedev,
doi:10.1088/1475-7516/2025/07/003
[arXiv:2503.14652 [hep-ph]].
   
   
\bibitem{Ema:2015dka}
Y.~Ema, R.~Jinno, K.~Mukaida and K.~Nakayama,
JCAP \textbf{05}, 038 (2015).

\bibitem{Bezrukov:2007ep}
F.~L.~Bezrukov and M.~Shaposhnikov,
Phys. Lett. B \textbf{659}, 703-706 (2008).

\bibitem{Enqvist:2015sua}
K.~Enqvist, S.~Nurmi, S.~Rusak and D.~Weir,
JCAP \textbf{02}, 057 (2016).
   
   
\bibitem{Cosme:2024ndc}
C.~Cosme, F.~Costa and O.~Lebedev,
JCAP \textbf{06}, 031 (2024).
   
   
\bibitem{Kolb:1990vq}
E.~W.~Kolb and M.~S.~Turner,
Front. Phys. \textbf{69}, 1-547 (1990).
   
   
\bibitem{Lebedev:2012sy}
O.~Lebedev and A.~Westphal,
Phys. Lett. B \textbf{719}, 415-418 (2013).
   
   
   
\bibitem{Laine:2016hma}
M.~Laine and A.~Vuorinen,
``Basics of Thermal Field Theory,''
Lect. Notes Phys. \textbf{925}, pp.1-281 (2016),
Springer, 2016.
   
   
\bibitem{Greene:1998nh}
P.~B.~Greene and L.~Kofman,
Phys. Lett. B \textbf{448}, 6-12 (1999).
   
     
\bibitem{Ichikawa:2008ne}
K.~Ichikawa, T.~Suyama, T.~Takahashi and M.~Yamaguchi,
Phys. Rev. D \textbf{78}, 063545 (2008).    
 
 
\bibitem{Dodelson:1993je}
S.~Dodelson and L.~M.~Widrow,
Phys. Rev. Lett. \textbf{72}, 17-20 (1994).
 
 
\bibitem{Hall:2009bx}
L.~J.~Hall, K.~Jedamzik, J.~March-Russell and S.~M.~West,
JHEP \textbf{03}, 080 (2010).  
   
   
\bibitem{Cosme:2023xpa}
C.~Cosme, F.~Costa and O.~Lebedev,
Phys. Rev. D \textbf{109}, no.7, 075038 (2024).
   
   
\bibitem{Planck:2018jri}
Y.~Akrami \textit{et al.} [Planck],
Astron. Astrophys. \textbf{641}, A10 (2020).
   
\bibitem{BICEP:2021xfz}
P.~A.~R.~Ade \textit{et al.} [BICEP and Keck],
Phys. Rev. Lett. \textbf{127}, no.15, 151301 (2021).
   
   
   
\bibitem{Lebedev:2022ljz}
O.~Lebedev and J.~H.~Yoon,
JCAP \textbf{07}, no.07, 001 (2022).
   
   
\bibitem{Kusenko:2006rh}
A.~Kusenko,
Phys. Rev. Lett. \textbf{97}, 241301 (2006).
   
\bibitem{Lebedev:2021xey}
O.~Lebedev,
Prog. Part. Nucl. Phys. \textbf{120}, 103881 (2021).
   
   
\bibitem{Greene:1997fu}
P.~B.~Greene, L.~Kofman, A.~D.~Linde and A.~A.~Starobinsky,
Phys. Rev. D \textbf{56}, 6175-6192 (1997).
   
   
   
\bibitem{Khlebnikov:1996mc}
S.~Y.~Khlebnikov and I.~I.~Tkachev,
Phys. Rev. Lett. \textbf{77}, 219-222 (1996).
   
   
\bibitem{DeRomeri:2020wng}
V.~De Romeri, D.~Karamitros, O.~Lebedev and T.~Toma,
JHEP \textbf{10}, 137 (2020).


\bibitem{Lebedev:2023uzp}
O.~Lebedev and T.~Toma,
JHEP \textbf{05}, 108 (2023).
   
   
     \bibitem{diff-eq}
      Valentin  Zaitsev and  Andrei  Polyanin,
      ``Handbook of Exact Solutions for Ordinary Differential Equations'', 
      Chapman and Hall/CRC, 2002.

   
   
\bibitem{Hannestad:2004px}
S.~Hannestad,
Phys. Rev. D \textbf{70}, 043506 (2004).

   
   \end{thebibliography}
\end{document}